\documentclass[a4paper,11pt,final]{article}

\usepackage{authblk}

\usepackage{graphics}
\usepackage{graphicx}
\usepackage{epsfig}
\usepackage{amssymb}
\usepackage{amsthm}

\usepackage{amsmath}
\date{}

\begin{document}


\title{Markovian nature, completeness, regularity and correlation properties of Generalized  Poisson-Kac processes}


\author[1]{Massimiliano Giona$^*$}
\author[2]{Antonio Brasiello}
\author[3]{Silvestro Crescitelli}
\affil[1]{Dipartimento di Ingegneria Chimica DICMA
Facolt\`{a} di Ingegneria, La Sapienza Universit\`{a} di Roma
via Eudossiana 18, 00184, Roma, Italy  \authorcr
$^*$  Email: massimiliano.giona@uniroma1.it}

\affil[2]{Dipartimento di Ingegneria Industriale
Universit\`{a} degli Studi di Salerno
via Giovanni Paolo II 132, 84084 Fisciano (SA), Italy}

\affil[3]{Dipartimento di Ingegneria Chimica,
 dei Materiali e della Produzione Industriale
Universit\`{a} degli Studi di Napoli ``Federico II''
piazzale Tecchio 80, 80125 Napoli, Italy}
\maketitle

\begin{abstract}
We analyze some basic issues associated with Generalized  Poisson-Kac (GPK) 
stochastic
processes, starting from the extended notion of  the Markovian condition.
The extended Markovian  nature of GPK processes
is established, and the implications of this property 
derived: the associated adjoint
formalism  for GPK processes is developed 
essentially in an  analogous way as for the Fokker-Planck operator 
associated with  Langevin equations driven by Wiener processes.
Subsequently, the regularity of trajectories
is addressed: the occurrence of fractality in the
realizations of GPK is a
 long-term
emergent property, and
 its implication in  thermodynamics is
discussed. The concept of completeness in the stochastic
description of GPK
is also introduced. Finally, some observations on
the role of correlation properties of noise sources
and their influence on 
 the dynamic properties of transport phenomena  are addressed, using a Wiener
model for comparison.
\end{abstract}


\section{Introduction}
\label{sec_1}

Stochastic models 
find a wide and fertile application in all the  fields
of physical science.
Notwithstanding their ubiquitous
application  in physics
and the huge literature existing on it,
some  basic  issues,  alternatives 
equation setting and, sometimes, misunderstandings
in the definition of basic properties  are still matter
of research and debate \cite{ottinger}.
For example, the choice of the most suitable stochastic calculus (Ito, Stratonovich, Klimontovich)
in physical applications is still matter of debate \cite{vankampen,stratoito1,stratoito2}, although Wong-Zakai theorem \cite{wongzakai1,wongzakai2} and the 
consistency of a thermodynamic description \cite{sekimoto} support
the Stratonovich approach (related to this debate see also \cite{klimo1}). 
Similarly, it is still
open the issue of the most appropriate extensions
of Brownian motion in a relativistic framework (different
model exist, the Relativistic Brownian Motion, the Relativistic Ornstein-Uhlenbeck Process) possessing similar steady-state distributions (the Juttner
distribution), but completely different dynamic properties \cite{dunkelrel}.
Sometimes, controversies arise due to the fact
the same diction and terminology is used to
indicate qualitative different stochastic phenomenologies, and {\em 
vice versa}
(as we shall see in the remainder of this article).

Therefore,
great attention should be posed in seeking the level of highest clarity
in assumptions and definitions, distinguishing
between fundamental results of general validity 
and phenomenologies deriving from  particular models, in the strive
of constructing a coherent picture of the physical reality upon stochastic
models.

This imperative is particularly compelling in contemporary research,
at least for two classes of reasons.  From one hand, experimental
research at micro- and nanoscales permits to obtain 
accurate measurements of the trajectories of single microsized
colloidal particles \cite{microparticle1,microparticle2},
and to analyze transport phenomena  (thermal problems) at extremely  short
time-scales (picosecond and attosecond physics) \cite{attosecond1,attosecond2,attosecond3}.
 On the other hand, 
new advancements of non-equilibrium physics
(fluctuation-dissipation theorems far from equilibrium \cite{fluctdiss1,fluctdiss2},
stochastic thermodynamics \cite{sekimoto}, extended thermodynamics \cite{extendedthermodynamic00,extendedthermodynamic0,extendedthermodynamic}),
 stimulate the research on more
general stochastic processes, and their connection with non-equilibrium
phenomena, capable of solving the shortcomings and limitations
of the existing theories.

This article deals with a class of stochastic processes, the
Generalized Poisson-Kac (GPK, for short) 
processes recently introduced in \cite{giona_gpk}, 
and their basic properties.
GPK processes, briefly reviewed in Section \ref{sec_2}, emerge 
as a generalization of the stochastic model proposed
by Mark Kac \cite{kac}, based on Poissonian fluctuations,
representing the stochastic  counterpart of  the one-dimensional
Cattaneo diffusion equation with memory \cite{cattaneo}.
Poisson-Kac processes solve the unpleasant feature of Wiener
processes, and Wiener-driven Langevin equation,
 of possessing  infinite propagation velocity.

In the physical literature, Poisson-Kac
processes are known and studied
under the dictions of ``dichotomous noise'', ``two-level bounded
noise'', and they have been used essentially
 as a simple and analytically
approachable model of colored noise, because of the exponential
decay in time of the correlation function \cite{dicho1,dicho2,dicho3,dicho4,dicho5,dicho5bis,dicho6}. Although a huge literature
has studied the dichotomous noise, the overwhelming  majority
of the existing contributions deal with
   one-dimensional spatial
models, with the exception of \cite{plykhin} and of  a series
of significant mathematical works by Kolesnik and coworkers
\cite{kolesnik1,kolesnik2,kolesnik3,kolesnik4} on two-dimensional systems
and their complex statistical description.

The importance of Poisson-Kac processes in physics goes far beyond their
``colored'' (in the meaning of {\em colored noise})  applications.
As originally observed by Gaveau et al. \cite{kacdirac},
and subsequently elaborated
by other authors \cite{dirac1,dirac2,dirac3}, Poisson-Kac processes provide a stochastic
interpretation to the relativistic Dirac equation for the free electron
(in the first paper \cite{kacdirac}, the 
one-dimensional Dirac equation is considered),
via analytic continuation of the temporal variable. Moreover,
under certain conditions,   referred to as the Kac limit, 
Poisson-Kac processes
 converge to a
Langevin equation driven by Wiener fluctuations \cite{kac,kacconvergence}.

These two properties, gathered together, are extremely appealing
in theoretical physics (field theories, matter-radiation interaction,
fluctuation theory at micro- and nanoscales). On the one hand,
all the classical results of stochastic Langevin theory
can be recovered (in the Kac limit). On the other hand, the finite
propagation velocity characterizing these processes provides a key
for the stochastic modeling of electromagnetic fluctuations
(zero-point energy, spontaneous emission and adsorption of
radiation, etc.). The latter property
is relevant in the  development of a stochastic field  theory
of field-matter interactions as the fundamental mechanism of
irreversibility and of the exotic properties
of matter at the atomic and subatomic scale grounded
on stochastic electrodynamics \cite{stochel1,stochel2,stochel3,stochel4}.

There is another field in which GPK process can be conveniently
used, namely {\em extended thermodynamics} \cite{extendedthermodynamic,
exttermo2,exttermo3,exttermo4}.
GPK processes provide a firm stochastic ground on the
generalization of the thermodynamic formalism referred
to as  {\em extended thermodynamics} of non-equilibrium processes
 \cite{extendedthermodynamic,
exttermo2,exttermo3}, pioneered by  Muller and  Ruggeri
and subsequently elaborated by many other authors,
especially Jou, Casas-Vazquez and Lebon \cite{exttermo3}.
Extended thermodynamic theories represent
a highly valuable contribution in generalizing
the classical theories of irreversible phenomena \cite{degroot}
essentially in two main directions: (i)  by including explicitly
thermodynamic fluxes in the representation
of the thermodynamic state functions in far-from-equilibrium conditions
and, (ii) in connecting this generalization with
the requirement of finite propagation velocity.
Nevertheless, the existing formulations of extended
thermodynamics  and the  related transport theories
still suffer some basic conceptual issues, that should
be addressed and solved in order to toughen and make fully
consistent the formal apparatus of these theories.

The theory, originated by the Grad 13-moment expansion
of the Boltzmann equation \cite{grad}
 is essentially based, as  mathematical
building block, on the multidimensional Cattaneo equation.
In space dimensions higher than one,
the Cattaneo equation is not probabilistically
consistent,  in the meaning that there is no
stochastic process for which the Cattaneo equation
is the evolution equation for its probability density function.
This property implies in turn that the higher-dimensional
Cattaneo equation
 does not preserve non-negativity \cite{cattaneonegative}, which
is highly unpleasant when this equation
is applied to describe the space-time 
evolution of molecular concentrations, absolute temperature
fields,
or probability density functions, which by
definition attain strictly non-negative values.

Recently, we have started a systematic
study of Poisson-Kac processes and of their higher-dimensional
extensions (GPK)  with the fundamental goals just
mentioned above \cite{giona2,giona3,giona_bc,giona5,giona6}.
In developing such a research program, one encounters some
delicate issues related to the very basic properties of Poisson-Kac 
and GPK
processes, that require a careful dissection in order to avoid
misunderstanding and misinterpretations.

One of this issues is  the Markovian
nature of these processes.
This  and related issues are analyzed in the present article. In point of fact,
starting from the understanding of   general concepts, 
new formalisms and novel results are developed.

This article is organized as follows. Section \ref{sec_2}
briefly reviews  GPK processes and  their
properties  that are useful in the remainder.
Section \ref{sec_3} discusses their Markovian nature, and
from the extended Markovian condition, the adjoint formalism
for GPK processes is developed, tracking the
analogy with the classical Kolmogorov theory of forward and
 backward Fokker-Planck equations.
Section \ref{sec_4} discusses the emerging fractality
of these processes versus their local (short-term regularity),
analyzing some of their implications  in
stochastic energetics (Section \ref{sec_5}).

Closely, related to these issues is the concept of
completeness of stochastic description, addressed in Section \ref{sec_6}.
Finally, \ref{sec_7} discussed the role of correlations in Poisson-Kac
processes.
Comparing a Poisson-Kac
dynamics with the evolution of a Langevin equation driven
by Wiener fluctuations,  possessing the same exponentially
decaying correlation function, it is shown that
the two model do possess radically different qualitative properties.
This observation can be further elaborated in order to describe
qualitatively different tunneling phenomenologies in stochastic dynamics.

\section{Generalized Poisson-Kac processes}
\label{sec_2}

The classical prototype for Poisson-Kac processes is represented by the
stochastic differential equation driven by Poisson noise \cite{kac},
that, in  one-dimensional spatial systems, attains the form 
\begin{equation}
d x(t) = v(x(t)) \, dt + b \, (-1)^{\chi(t)} \, d t 
\label{eq2_1}
\end{equation}
where $v(x)$ is a deterministic velocity field, $b>0$ is the
characteristic velocity of the stochastic fluctuations, and
$\chi(t)$ is a Poisson process possessing transition rate $a>0$.
The stochastic contribution $(-1)^{\chi(t)}$ switches between $+1$ and
$-1$, and this is the reason why this model is often referred to
as ``dichotomous noise'' or ``two-level noise'' \cite{dicho1,dicho2,dicho3,dicho4,dicho5}.
The initial condition for $\chi(t)$ at $t=0$ is usually set
$\chi(0)=0$ with probability 1, as for classical Poisson processes. Alternatively,
 in order to
avoid any form of  bias at $t=0$, it can be set 
$\chi(0)=0$ and $\chi(0)=1$ with 
$\mbox{Prob}[\chi(0)=0]=\mbox{Prob}[\chi(0)=1]=1/2$.

Indicate with $X(t)$ the stochastic process at time $t$ originated
by eq. (\ref{eq2_1}), and let 
\begin{equation}
p^\pm(x,t) \, d x= \mbox{Prob}[ X(t) \in (x,x+dx), \; (-1)^{\chi(t)}= \pm 1 \}
\label{eq2_2}
\end{equation}
be the associated partial probability densities.
The partial probabilities $p^{\pm}(x,t)$  satisfy the
balance equations \cite{kac,giona3}
\begin{equation}
\partial_t p^{\pm}(x,t) = - \partial_x \left [ v(x) \, p^\pm(x,t) \right ]
\mp b \, \partial_x p^\pm(x,t) \mp a \,  \left [p^{+}(x,t) - 
p^{-}(x,t)  \right ]
\label{eq2_3}
\end{equation}
In the absence of the deterministic
contribution $v(x)$, 
the solutions of these equations correspond to waves propagating
in the forward ($p^+$), and backward ($p^-$) $x$-direction,
with mutual recombination controlled
by the exponential statistics of the switching times. For
this reason, these functions are also referred to
as partial probability waves of the Poisson-Kac process.
The  overall
probability density function (pdf) $p(x,t)$ for $X(t)$ at time $t$
is  thus given by 
\begin{equation}
p(x,t)=p^+(x,t)+p^-(x,t)
\label{eq2_3a}
\end{equation}

In the limit for $a,b \rightarrow \infty$, keeping fixed and
constant the ratio $b^2/2 a=D$ (where $D$ has the physical
dimension of a diffusion constant), the pdf $p(x,t)$
becomes the solution of a classical parabolic
advection-diffusion equation 
\begin{equation}
\partial_t p(x,t) =  - \partial_x \left [ v(x)
\, p(x,t) \right ] + D \, \partial_x^2 p(x,t)
\label{eq2_3b}
\end{equation}
 that corresponds to
the forward Fokker-Planck equation associated with the
Langevin equation
\begin{equation}
d x(t) = v(x(t)) \, d t + \sqrt{2 \, D} \, d w(t)
\label{eq2_3add}
\end{equation}
where $d w(t)$ is the increment in the interval $d t$ of a one-dimensional
Wiener process. Eq. (\ref{eq2_3add})
 represents the Kac limit for eq. (\ref{eq2_1})
while eq. (\ref{eq2_3b})
 expresses the  statistical convergence of the Poisson-Kac process
towards the Fokker-Planck equation associated
with the  Langevin model (\ref{eq2_3add}). For further
mathematical details on the convergence see \cite{kacconvergence}.

\subsection{Generalized Poisson-Kac processes}
\label{sec_2_1}

The Generalized Poisson-Kac processes represent a generalization of eq. 
(\ref{eq2_1}), particularly suited for modeling stochastic systems
in spatial dimensions higher than one \cite{giona_gpk}. 

The starting point is the $N$-state finite Poisson process $\chi_N(t)$,
which is a stochastic process attaining,
  at any $t>0$, $N$ distinct values $\chi_N(t)=1,2,\dots,N$.  As the
classical Poisson process, it is a stationary, memoryless and ordinary
process, so that in a time interval $\Delta t$, the transition probabilities
from $\chi_N(t)=\alpha$ to $\chi_N(t+\Delta t)=\beta$, $\alpha,\beta=1,\dots, N$, are given by
\begin{equation}
T_{\alpha \mapsto \beta}(\Delta t) = \lambda_\alpha \, A_{\beta,\alpha} \,
\Delta t
+ o(\Delta t) 
\label{eq2_5}
\end{equation}
and
\begin{equation}
T_{\alpha \mapsto \alpha}(\Delta t) =
\left [ 1- \lambda_\alpha \, \sum_{\beta=1}^N A_{\beta,\alpha} \right ]
\, \Delta t
+ o(\Delta t) 
\label{eq2_6}
\end{equation}
where $\lambda_\alpha>0$ are the transition rates and ${\bf A}=(A_{\alpha,\beta}
)_{\alpha,\beta=1}^N$ is a left-stochastic matrix,
\begin{equation}
A_{\beta,\alpha} \geq 0 \, , \qquad \sum_{\beta=1}^N A_{\beta,\alpha}=1 
\;\;\; \alpha=1,\dots,N
\label{eq2_7}
\end{equation}

Let $\{ {\bf b}_\alpha \}_{\alpha=1}^N$ be a system of $N$ constant
vectors of ${\mathbb R}^d$, $d=1,2,\dots$, fulfilling  the
non-biasing condition $\sum_{\alpha=1}^N {\bf b}_\alpha =0$.

A {\em Generalized Poisson Kac} process ${\bf X}(t)$ (henceforth GPK) in
${\mathbb R}^d$, in the presence of a deterministic velocity field
${\bf v}({\bf x})$, is defined by the following
stochastic differential equation for its realizations ${\bf x}(t)$
\begin{equation}
d {\bf x}(t) = \left [ {\bf v}({\bf x}(t)) + {\bf b}(\chi_N(t))
\right ] \, d t 
\label{eq2_8}
\end{equation}
where 
\begin{equation}
{\bf b}(\chi_N(t)=\alpha)={\bf b}_\alpha
\label{eq2_8a}
\end{equation}
 and $\chi_N(t)$
is a $N$-state finite Poisson process.
Taking into account the properties of $\chi_N(t)$, essentially
the fact that it takes only $N$ distinct values, the
GPK is fully described by the the conditional
probability density functions
 $p_{\alpha,\beta}({\bf x},t /{\bf y},t_0)$,
$\alpha,\beta=1,\dots,N$. 
Indicating with $d {\bf x}$  the volume element in ${\mathbb R}^d$,
$p_{\alpha,\beta}({\bf x},t /{\bf y},t_0) \, d {\bf x}$,
represent the probabilities that ${\bf X}(t) \in ({\bf x},{\bf x}+ d{\bf x})$
and $\chi_N(t)=\alpha$ conditional to ${\bf X}(t_0)={\bf y}$
and $\chi_N(t_0)=\beta$, where $t_0 <t$,
\begin{eqnarray}
p_{\alpha,\beta}({\bf x},t / {\bf y}, t_0) \, d {\bf x}
& = & \mbox{Prob} \left [  \, {\bf X}(t) \in ({\bf x} , {\bf x}+d {\bf x}) \,  , \; \;
\chi_N(t)=\alpha \right . \nonumber \\
    & / &  \left .  {\bf X}(t_0)={\bf y} \,, \; \; \chi_N(t_0)=\beta \, 
\right ] 
\label{eq2_8b}
\end{eqnarray}
Dropping for notational simplicity the indication about the
dependence on the initial state ${\bf y}$ at time $t_0$,
the conditional probabilities $p_{\alpha,\beta}({\bf x},t)=
p_{\alpha,\beta}({\bf x},t / {\bf y},\beta)$  satisfy the wave-equations
\begin{eqnarray}
\partial_t p_{\alpha,\beta}({\bf x}, t) & = & -
\nabla_x \cdot \left [ ({\bf v}({\bf x})+ {\bf b}_\alpha) \,
p_{\alpha,\beta}({\bf x},t) \right ]- \lambda_\alpha \, p_{\alpha,\beta}({\bf x},t ) \nonumber \\
& + & \sum_{\gamma=1}^N \lambda_\gamma \, A_{\alpha,\gamma} \, p_{\gamma,
\beta}({\bf x},t) = {\mathcal L}_{x, \alpha} [p_{\alpha,\beta}({\bf x},t); \{
p_{\gamma,\beta}\}_{\gamma=1}^N]
\label{eq2_9}
\end{eqnarray}
where 
$\nabla_x$ indicates the nabla-operator
with respect to the ${\bf x}$-variables. 
The second argument, namely $\{ p_{\gamma,\beta}\}_{\gamma=1}^N$, of
the operator ${\mathcal L}_{x,\alpha}$, indicates that
this operator depends  on  the whole
system of probability densities 
$p_{1,\beta}({\bf x},t),\dots,p_{N,\beta}({\bf x},t)$, 
evaluated at $({\bf x},t)$ corresponding to space-time point of the
first argument of ${\mathcal L}_{x,\alpha}$,
keeping fixed  the second index, namely $\beta$,
 of the conditional probabilities.
It follows from the 
structure of eq. (\ref{eq2_9}) that the partial  probability waves
\begin{equation}
\overline{p}_\alpha({\bf x},t/{\bf y}, t_0)= 
\sum_{\beta=1}^N p_{\alpha,\beta}({\bf x},t/ {\bf y}, t_0)
\label{eq2_9a}
\end{equation}
 defined from
the conditional probabilities $p_{\alpha,\beta}({\bf x},t/{\bf y}, t_0)$,
 averaging out the information on the initial state of
the process $\chi_N(t_0)$, i.e., summing over $\beta$,
 satisfy the same wave equation, namely,
\begin{equation}
\partial_t \overline{p}_\alpha({\bf x},t/{\bf y},t_0) = {\mathcal L}_{x,\alpha}[\overline{p}_\alpha({\bf x},t/{\bf y}, t_0);\{\overline{p}_\gamma \}_{\gamma=1}^N] 
\label{eq2_10}
\end{equation}

\subsection{A two-dimensional model}
\label{sec2_2}

Consider the two-dimensional Poisson-Kac process in the absence of
deterministic biasing field,
\begin{equation}
\left \{
\begin{array}{l}
d x(t) = b \, (-1)^{\chi_1(t)} \, dt \\
d y (t) = b \, (-1)^{\chi_2(t)} \, dt  
\end{array}
\right .
\label{eq2_11}
\end{equation}
where $\chi_1(t)$, $\chi_2(t)$ are two independent Poisson processes 
possessing the same statistical properties, namely the same transition rate $a$. This process can be viewed as a GPK with $N=4$. In this case, all the
$\lambda_\alpha$, $\alpha=1,\dots,4$ coincide and are equal to $2 \, a$.
The vectors ${\bf b}_\alpha$ are given by:
${\bf b}_1=b  \, (1,1)$, ${\bf b}_2= b \, (-1,1)$, ${\bf b}_3=b \, (1,-1)$,
${\bf b}_4=b \, (-1,1)$, where state ``$1$''
 corresponds to $(-1)^{\chi_1(t)}=1$,
$(-1)^{\chi_2(t)}=1$, state ``2'' to  $(-1)^{\chi_1(t)}=-1$
, $(-1)^{\chi_2(t)}=1$, state ``3'' to $(-1)^{\chi_1(t)}=1$
, $(-1)^{\chi_2(t)}=-1$, state ``4'' to
$(-1)^{\chi_1(t)}=-1$, $(-1)^{\chi_2(t)}=-1$. The transition
probability matrix ${\bf A}$ is given by
\begin{equation}
{\bf A}= \left (
\begin{array}{llll}
0 & 1/2 & 1/2 & 0 \\
1/2 & 0 & 0 & 1/2 \\
0 & 1/2 & 1/2 & 0 \\
1/2 & 0 & 0 & 1/2 
\end{array}
\right ) 
\label{eq2_12}
\end{equation}
The probability of occurrence of the transitions
$(-1,-1) \mapsto (1,1)$ or $(1,1) \mapsto (-1,-1)$ in the
time interval $\Delta t$ is order of ${\mathcal O}(\Delta t^2)$,
and therefore they are negligible in the limit for $\Delta t \rightarrow 0$.
Observe that in this case ${\bf A}$ is a doubly stochastic matrix.
We will use this example, in Section \ref{sec_4} for addressing
the regularity of GPK trajectories and the emergent character of their
fractal properties.

Similarly, every Poisson-Kac process involving $M$ independent
Poisson processes can be viewed as a GPK process involving $N=2^M$
states. The reverse  property is  not  necessarily true.

\section{Markovian condition and adjoint theory for GPK} 
\label{sec_3}

The Markovian nature of dichotomous noise processes (and, {\em a fortiori},
of GPK) is a controversial issue in the Literature, as some authors
have claimed their non-Markovian character \cite{nonmarkov1,nonmarkov2}, 
while others affirm
their Markovian nature \cite{kolesnik1,markov1}.
This issue is essentially of semantical nature, albeit it disclosures
relevant consequences in the development of the  GPK theory.

In some sense - paraphrasing Plato - the Markovian nature of a process
``lies in the eye of the beholder'', as it depends essentially  on
the way it is defined. 
Intuitively, the Markovian character of a process accounts for its
memoryless property, in that the {\em complete statistical information}
on the process at any given time $t_0$
is fully sufficient to achieve a
complete statistical knowledge of the process at any later time $t>t_0$.

This is the essence of the Markov condition, albeit the last sentence 
disclosures a delicate issue, 
related to the formal and exact meaning of the concept of
 ``complete statistical
description''.
A typical example of this apparent ambiguity is represented
by the concept of Markovian embedding \cite{markovianembedding}.

Consider a Langevin equation of the form
\begin{equation}
d x(t) = v(x(t), w(t))\, d t + a(x(t),w(t)) \, d w(t) 
\label{eq3_1}
\end{equation}
where $w(t)$ is a one-dimensional Wiener process, and the
coefficients $v(x,w)$, $a(x,w)$ depend not solely on the
process $X(t)$ at time $t$,
 but also on the stochastic  perturbation $W(t)$
at the same time $t$.
This process is strictly-speaking, non Markovian, as the probability 
density function $p(x,t)$ does not satisfy the Markovian condition
\begin{equation}
p(x,t+\tau / x_0, t_0)= \int p(x, t+\tau / y, t) \, p(y,t / x_0, t_0) \, d y 
\label{eq3_2}
\end{equation}
where $t_0< t$ and $\tau>0$. Next, 
 introduce the auxiliary process $Y(t)=W(t)$, i.e.,
\begin{equation}
d y(t) = d w(t) 
\label{eq3_3}
\end{equation}
with $y(0)=0$.
With this extension, the vector-valued stochastic process
${\bf X}(t)=(X(t),Y(t))$ the dynamics
of which is expressed by eq. (\ref{eq3_3}) and by
eq. (\ref{eq3_1}), which now
can be  rewritten as 
\begin{equation}
d x(t)=v(x(t),y(t)) \, d t+ a(x(t),y(t)) \, d w(t)
\label{eq3_3add}
\end{equation}
 is Markovian. This is the essence
of the technique of Markovian embedding \cite{markovianembedding}, 
and of the concept
of generalized  (hidden) 
Markov processes \cite{generalizedmarkov1,generalizedmarkov2}, 
which is conceptually similar
either to the concept of suspension for
deterministic non-autonomous systems,
 in order to transform them in autonomous ones
by adding an additional phase-space coordinate 
 homeomorphic to time \cite{guckenheimer},
and to the deterministic embedding of attractors due to Takens \cite{takens},
in order to reconstruct a chaotic attractor from a single time series.

Keeping in mind  the essential meaning of the
 Markov property, namely that the
statistical knowledge of a process at time $t$ 
is fully determined by its statistical
information at any  preceding time $t_0<t$, and it is immaterial of its
history preceding $t_0$, a Markov condition can be stated also for GPK,
bearing in mind that all the statistical  information needed
to describe the system is contained in the conditional probability
densities
$\{ p_{\alpha,\beta}({\bf x},t/{\bf y}, t_0) \}_{\alpha,\beta=1}^N$
that keep also track of the state of the $N$-state finite Poisson process 
generating the randomness in the system.

In order to avoid confusion with the concept of generalized Markovian
processes, it is convenient to introduce the definition
of {\em extended Markovian processes}.

A stochastic vector-valued process $ {\bf Z}(t)$ is said to be an
 extended Markovian process
if: (i) there exist $M>0$ 
non-negative functions $\pi_k({\bf z},t/{\bf z}_0,t_0)$ 
such that the
conditional probability density
 $p({\bf z},t / {\bf z}_0,t_0)$ of finding the
state of the process in ${\bf z}$ at time $t$, given the initial state ${\bf z}_0$ at 
time $t_0<t$
is given by
\begin{equation}
p({\bf z},t/ {\bf z}_0, t_0) = \sum_{k=1}^M \pi_k({\bf z},t/ {\bf z}_0, t_0) 
\label{eq3_4}
\end{equation}
and (ii) there exits $M^3$ non-negative constants $C_k^{p,q}$ such that
the functions $ \pi_k({\bf z},t/ {\bf z}_0, t_0) $ satisfy the recurrence equation
for any ${\bf z}, \, {\bf z}_0 $ and $t_0<t_1<t$,
\begin{equation}
\pi_k({\bf z},t / {\bf z}_0, t_0) = \sum_{p,q=1}^M C_k^{p,q} \int 
\pi_p({\bf z},t / {\bf y}, t_1) \, \pi_q({\bf y},t_1 / {\bf z}_0, t_0)
\, d {\bf y} 
\label{eq3_5}
\end{equation}

Horsthemke and Lefevre  \cite{dicho2}
have proved that the extended Markovian property holds
for the one-dimensional  Poisson-Kac model eq. (\ref{eq2_1}) driven by
dichotomous Poisson noise. In the setting of GPK processes,
 the Horsthemke-Lefevre condition
becomes (for $t_1 <t_2 < t_3)$,
\begin{equation}
p_{\alpha,\beta}({\bf x},t_3 /{\bf  z}, t_1) = \sum_{\gamma=1}^N
\int p_{\alpha,\gamma}({\bf x},t_3/{\bf y},t_2) \, p_{\gamma, \beta}({\bf y},t_2/{\bf z},t_1) \, d {\bf y} 
\label{eq3_6}
\end{equation}
$\alpha,\beta=1,\dots,N$.
Enforcing the transition
 properties of a $N$-time finite Poisson process,
 it is straightforward to
derive the balance equation eq. (\ref{eq2_9}) from the extended
 Markovian relation
 (\ref{eq3_6}). The technique is essentially analogous to that
used in \cite{dicho2} and its is not repeated here.

In  point of fact, eq. (\ref{eq3_6}) corresponds to
the extended Markov property defined by eq. (\ref{eq3_5})
where  the conditional
probability densities $\{ p_{\alpha,\beta} \}_{\alpha,\beta=1}^N$  play
the role of 
the $M=N^2$ auxiliary functions $\{\pi_k\}_{k=1}^M$.
Eq. (\ref{eq3_6}) can be thus  rewritten in the form of eq. (\ref{eq3_5}), i.e.,
\begin{equation}
p_{\alpha,\beta}({\bf x},t_3 / {\bf z}, t_1) =
\sum_{\gamma,\epsilon,\zeta,\xi=1}^N C_{(\alpha,\beta)}^{(\gamma,\epsilon), (\zeta,\xi)} \int p_{\gamma,\epsilon}({\bf x},t_3/{\bf y},t_2) \, p_{\zeta,\xi}({\bf y},t_2 /{\bf z},t_1) \, d {\bf y} 
\label{eq3_7}
\end{equation}
by introducing the coefficient tensor 
$C_{(\alpha,\beta)}^{(\gamma,\epsilon), (\zeta,\xi)}= 
\delta_\alpha^\gamma \, \delta_\beta^\xi \, \delta^{\epsilon,\zeta}$,
where $\delta_\alpha^\gamma$, $\delta^{\epsilon,\zeta}$ are the Kronecker
symbols, written in countervariant and mixed way purely because
 of notational 
index consistency.

\subsection{Adjoint formalism and first transit-time statistics}
\label{sec_3_1}

A mathematical definition is useful in physics if,
starting from it, it is possible to derive
physical consequences on the process under investigation.
This is the case of the extended Markov property that permits to
derive the adjoint formalism for GPK processes and the first-transit
description, conceptually analogous to that of
strictly Markovian processes associated with Langevin equations 
driven by Wiener fluctuations.

Consider eqs. (\ref{eq3_6}), where the probabilities
$p_{\alpha,\beta}({\bf x},t_3/{\bf z},t_1)$ satisfies the
forward evolution equations (\ref{eq2_9}), below  rewritten   for
convenience
\begin{equation}
\partial_{t_2} p_{\gamma,\beta}({\bf y}, t_2 / {\bf z}, t_1)
= {\mathcal L}_{y,\gamma}[p_{\gamma,\beta}({\bf y}, t_2 / {\bf z}, t_1) ; \{p_{\xi,\beta}\}_{\xi=1}^N] 
\label{eq3_8}
\end{equation}

Summing in eq. (\ref{eq3_6}) over the indexes $\alpha,\, \beta $,
 and integrating
over ${\bf x}$, the identity follows
\begin{equation}
1= \sum_{\alpha,\beta,\gamma=1}^N \int d {\bf x} \int d {\bf y} \,
 p_{\alpha,\gamma}({\bf x},t_3/{\bf y},t_2) \, p_{\gamma, \beta}({\bf y}, t_2 / {\bf z},t_1) 
\label{eq3_9}
\end{equation}
Taking the derivative with respect to $t_2$ and enforcing the
forward evolution equation (\ref{eq3_8}) for
 $p_{\gamma, \beta}({\bf y}, t_2 / {\bf z},t_1)$,
 one obtains
\begin{eqnarray}
0 & = & \sum_{\alpha,\beta,\gamma=1}^N \int d {\bf x} \int d {\bf y}
\left [
 p_{\gamma, \beta}({\bf y},t_2 /{\bf z},t_1) \,
 \partial_{t_2} p_{\alpha,\gamma}({\bf x},t_3/{\bf y},t_2)  \right .
\nonumber \\
& + & \left .  p_{\alpha,\gamma}({\bf x},t_3/{\bf y},t_2) \, {\mathcal L}_{y,\gamma} [ p_{\gamma, \beta}({\bf y},t_2/ {\bf z},t_1); \{p_{\xi,\beta} \}_{\xi=1}^N] \right ] 
\label{eq3_10}
\end{eqnarray}
Consider the  space $L^2_N({\mathbb R}^d)$ of $N$-dimensional
vector-valued square summable functions in ${\mathbb R}^d$
attaining real values as, in our case, probability densities are involved. 
This is a Hilbert  space equipped with the inner product
$( \cdot, \cdot)_{L^2_N}$: if ${\bf f}({\bf x})=(f_1({\bf x}),\dots,f_N({\bf x}))$, ${\bf g}({\bf x})=(g_1({\bf x}),\dots,g_N({\bf x}))$ belong
to $L^2_N({\mathbb R}^d)$, then
\begin{equation}
({\bf f},{\bf g})_{L^2_N} = \sum_{\gamma=1}^N \int  f_\gamma({\bf x})
\, g_\gamma({\bf x}) \, d {\bf x} 
\label{eq3_11}
\end{equation}
In eq. (\ref{eq3_10}), the integral over ${\bf y}$ and the sum
over $\gamma$ can be viewed as a scalar product in 
${L^2_N}({\mathbb R}^d)$ of the   transforms of the two 
functions ${\bf f}({\bf y},t_2)$ and
${\bf g}({\bf y},t_2)$, the entries of which are   given by 
\begin{equation}
f_\gamma({\bf y},t_2) = \sum_{\alpha=1}^N
p_{\alpha,\gamma}({\bf x},t_3 / {\bf y}, t_2) 
\label{eq3_12}
\end{equation}
and
\begin{equation}
g_\gamma({\bf y},t_2) = \sum_{\beta=1}^N p_{\gamma,\beta}({\bf y},t_2 /{\bf z}, t_1) 
\label{eq3_13}
\end{equation}
$\gamma=1,\dots,N$,
upon the action of the differential operators $\partial_{t_2}$ and $\{
 {\mathcal L}_{y,\gamma} \}_{\gamma=1}^N $
(the dependence on ${\bf x}$, $t_3$, ${\bf z}$, $t_1$ has been
omitted for notational simplicity), respectively.
Therefore, eq. (\ref{eq3_10}) can be expressed as 
\begin{equation}
0 = \int d {\bf x} \left [ (\partial_{t_2} {\bf f}, {\bf g} )_{L^2_N}
+ ({\bf f}, \widehat{{ \mathcal L}}_{y}[{\bf g}] )_{L^2_N} \right ]
\label{eq3_14}
\end{equation} 
where $\widehat{{\mathcal L}}_{y}$ is the vector-valued extension of the
$N$ differential linear operators ${\mathcal L}_{y,\alpha}$
\begin{equation}
\widehat{{ \mathcal L}}_{y} =
\left (
\begin{array}{c}
{\mathcal L}_{y,1} \\
{\mathcal L}_{y,2} \\
\cdot \\
\cdot \\
{\mathcal L}_{y,N}
\end{array}
\right ) \, .
\label{eq3_15}
\end{equation}
Let $\widehat {{\mathcal L}}_y^+$ be the adjoint of 
 $\widehat{{ \mathcal L}}_{y}$, which is simply the vector-valued
extension of the
adjoints ${\mathcal L}_{y,\alpha}^+$ of the
operators ${\mathcal L}_{y,\alpha}$,
\begin{equation}
\widehat{{ \mathcal L}}_{y}^+ =
\left (
\begin{array}{c}
{\mathcal L}_{y,1}^+ \\
{\mathcal L}_{y,2}^+ \\
\cdot \\
\cdot \\
{\mathcal L}_{y,N}^+
\end{array}
\right ) 
\label{eq3_16}
\end{equation}
Eq. (\ref{eq3_14}) thus becomes
\begin{equation}
0  =  \int d {\bf x} 
\left [ (\partial_{t_2} {\bf f}, {\bf g} )_{L^2_N}
+ (\widehat{ { \mathcal L}}_{y}^+[{\bf f}], {\bf g} )_{L^2_N} \right ]
= \int d {\bf x}  \,  (\partial_{t_2} {\bf f} + \widehat{ { \mathcal L}}_{y}^+[{\bf f}], {\bf g} )_{L^2_N} 
\label{eq3_17}
\end{equation}
and since it should be valid for any ${\bf g}$, it implies that
\begin{equation}
\partial_{t_2} {\bf f} + \widehat{ { \mathcal L}}_{y}^+[{\bf f}]=0 
\label{eq3_18}
\end{equation}
The entries of  the vector-valued
function ${\bf f}({\bf y},t_2)$ are,  by definition eq. (\ref{eq3_12}),
the backward-in-time partial probability  waves
\begin{equation}
f_\gamma({\bf y},t_2)=
 \sum_{\alpha=1}^N
p_{\alpha,\gamma}({\bf x},t_3 / {\bf y}, t_2) 
=
\overline{p}_\gamma^{(b)}({\bf x},t_3 / {\bf y},t_2) 
\label{eq3_19}
\end{equation}
(the summation involves the first inder $\alpha$ of $p_{\alpha,\gamma}$),
so that  eq. (\ref{eq3_18}) can be expressed as
\begin{equation}
\partial_{t_2}  \overline{p}_\gamma^{(b)}({\bf x},t_3 / {\bf y},t_2)
 + { \mathcal L}_{y,\gamma}^+[ \overline{p}_\gamma^{(b)}({\bf x},t_3 / {\bf y},t_2);
\{ \overline{p}_\beta^{(b)}\}_{\beta=1}^N]=0 
\label{eq3_20}
\end{equation}
which is the evolution equation, reversed in time, for the backward-in-time
partial probability waves. Due to the linearity of the definition
of the partial waves  $\overline{p}_\gamma^{(b)}({\bf x},t_3 / {\bf y},t_2)$
eq. (\ref{eq3_19}), and the independence of 
the equation (\ref{eq3_20}) of the final state $\alpha$, the same equation applies to
each individual term $p_{\alpha,\gamma}({\bf x},t_3/{\bf y},t_2)$,
namely
\begin{equation}
\partial_{t_2}  p_{\alpha,\gamma}({\bf x},t_3 / {\bf y},t_2)
 + { \mathcal L}_{y,\gamma}^+[ p_{\alpha,\gamma}({\bf x},t_3 / {\bf y},t_2);
\{p_{\alpha,\beta} \}_{\beta=1}^N]=0 
\label{eq3_21}
\end{equation}
where by eq. (\ref{eq2_9}) the adjoint operator ${ \mathcal L}_{y,\gamma}^+$,
$\gamma=1,\dots,N$,
are given by
\begin{eqnarray}
{\mathcal L}_{x, \gamma}^+ [p_{\alpha,\gamma}({\bf x},t); \{p_{\alpha,\beta}\}_{\beta=1}^N]
& = &
 ({\bf v}({\bf x})+ {\bf b}_\gamma )  \cdot \nabla_x  \,
p_{\alpha,\gamma}({\bf x},t) - \lambda_\gamma \, p_{\alpha,\gamma}({\bf x},t ) \nonumber \\
& + & \sum_{\beta=1}^N \lambda_\gamma \, A_{\beta,\gamma} \, p_{\alpha,
\beta}({\bf x},t) 
\label{eq3_22}
\end{eqnarray}
where we have set, for notational
simplicity, $p_{\alpha,\gamma}({\bf x},t)=p_{\alpha,\gamma}({\bf z},\tau/{\bf x},t)$, for generic ${\bf z}$ and $\tau$.

\subsection{First transit-time  statistics}
\label{sec_3_2}

The analysis of first transit times for dichotomous noise has been
developed in several articles,  almost exclusively considering the case
of one-dimensional spatial problems \cite{nonmarkov1,transit1,transit2,transit3,transit4,transit5,transit6,transit7}. The general formulation
of this  analysis does
not exists for GPK, and consequently it is developed below in the
most general setting.

Consider the first transit-time probability density functions.
For GPK processes, $N$ such probability densities  can be defined
$\widehat{F}_\beta({\bf x}/{\bf y},t)$, $\beta=1,\dots,N$, where
 $\widehat{F}_\beta({\bf x}/{\bf y},t)$
is the probability density function
 for the first transit-time through ${\bf x}$, given the initial
state  ${\bf y}$ at time $t=0$  and $\chi_N(0)=\beta$.
The definition of $\widehat{F}_\beta({\bf x}/{\bf y},t)$
implicitly assume that, starting from ${\bf y}$ at time $t=0$,
there is a trajectory of the process that reaches ${\bf x}$
in finite time. This is in general not true for all ${\bf x}$
and ${\bf y}$, in the presence of deterministic biasing field,
as the analysis developed in \cite{giona3} illustrates.
Under the above hypothesis,
 by  definition
\begin{equation}
\int_0^\infty \widehat{F}_\beta({\bf x}/{\bf y},t) \, d t = 1
\label{eq3_23}
\end{equation}
The density function $\widehat{F}_\beta({\bf x}/{\bf y},t)$ 
can be dissected also with
respect to the final state  of the
process $\chi_N(t)$  at the transit time $t$
 as
\begin{equation}
\widehat{F}_\beta({\bf x}/{\bf y},t)= \sum_{\alpha=1}^N
F_{\alpha,\beta}({\bf x} / {\bf y}, t) 
\label{eq3_24}
\end{equation}
where $F_{\alpha,\beta}({\bf x} / {\bf y}, t)$ corresponds to the
transit-time probability density of passing through ${\bf x}$
 at time $t$ with $\chi_N(t)=\alpha$, starting  from ${\bf y}$ at time $t=0$,
given that  $\chi_N(0)=\beta$.

The introduction of these quantities permits to derive easily
an equation for $\widehat{F}_\beta({\bf x}/{\bf y},t)$, using 
essentially the same strategy adopted in the strictly Markovian case.
Indeed,  the conditional probabilities 
$p_{\alpha,\beta}({\bf x}, t/{\bf y}, 0)$ can be expressed in terms
of $F_{\alpha,\beta}({\bf x}/{\bf y}, t)$ for $t>0$ as
\begin{equation}
p_{\alpha,\beta}({\bf x}, t/{\bf y}, 0)= \sum_{\gamma=1}^N
\int_0^t F_{\gamma,\beta}({\bf x} /{\bf y}, t-\tau) \,
p_{\alpha,\gamma}({\bf x}, \tau / {\bf x}, 0) \, d \tau
\label{eq3_25}
\end{equation}
$\alpha,\beta=1,\dots,N$.
Eqs. (\ref{eq3_25}) correspond to the fact that  the
probability $p_{\alpha,\beta}({\bf x}, t/{\bf y}, 0)$ 
is the integral over time $\tau$, from $\tau=0$ to the current time $\tau=t$,
and the sum over all the possible states $\gamma$,  
from $1$ to $N$,
of the probability  of arriving for the first time 
in ${\bf x}$, starting from ${\bf y}$ and
$\chi_N(0)=\beta$ at  time $t-\tau$ with $\chi_N(t-\tau)=\gamma$,
which is $F_{\gamma,\beta}({\bf x}/{\bf y}, t-\tau)$,
times the probability   
that in the remaining time $\tau$ the system will return to
${\bf x}$, i.e., $p_{\alpha,\gamma}({\bf x}, \tau / {\bf x}, 0)$.

But,
\begin{equation}
p_{\alpha,\beta}({\bf x},t/{\bf y},0) = p_{\alpha,\beta}({\bf x},0 / {\bf y}, -t) 
\label{eq3_26}
\end{equation}
and, because of eq.  (\ref{eq3_21}), this  quantity satisfies the
backward equation
\begin{equation}
\partial_t  p_{\alpha,\beta}({\bf x},t/{\bf y},0) -
{\mathcal L}_{y,\beta}^+ [ p_{\alpha,\beta}({\bf x},t/{\bf y},0); \{p_{\alpha,\xi} \}_{\xi=1}^N]=0 
\label{eq3_27}
\end{equation}
Consequently, the function at the r.h.s. of eq. (\ref{eq3_25})
satisfies eq. (\ref{eq3_27}) with respect to ${\bf y}$, 
and its direct substitution into
the latter equation provides the expression
\begin{eqnarray}
\sum_{\gamma=1}^N F_{\gamma,\beta}({\bf x} /{\bf y},0) \,  p_{\alpha,\gamma}({\bf x},t /{\bf y}, 0)   + \sum_{\gamma=1}^N \int_0^t \left \{
\partial_t F_{\gamma,\beta}({\bf x } / {\bf y}, t-\tau) \right .
\nonumber \\
 -  \left . 
 {\mathcal L}_{y,\beta}^+[F_{\gamma,\beta}({\bf x } / {\bf y}, t-\tau);
\{ F_{\gamma,\xi} \}_{\xi=1}^N
] \right \} \,
  p_{\alpha,\gamma}({\bf x}, \tau /{\bf x}, 0 ) \, d \tau =0 
\label{eq3_28}
\end{eqnarray}
For any ${\bf x} \neq {\bf y}$, $F_{\gamma,\beta}({\bf x} /{\bf y},0)=0$,
and since eq. (\ref{eq3_28}) should be valid for any ${\bf x}$ , it implies
that the integrand under parenthesis should be identically
vanishing. This returns the equation
for $F_{\gamma,\beta}({\bf x} / {\bf y},t)$ as a function of $t$ and
of the initial state ${\bf y}$,
\begin{equation}
\partial_t F_{\gamma,\beta}({\bf x} /{\bf y},t) =
{\mathcal L}_{y,\beta}^+[F_{\gamma,\beta}({\bf x} /{\bf y},t);\{F_{\gamma,\xi} \}_{\xi=1}^N] 
\label{eq3_29}
\end{equation}
which is a backward equation governed by the adjoint
 operator ${\mathcal L}_{y,\beta}^+$ of the initial state of the
 process $\chi_N(0)$, i.e., $\beta$.
Summing over $\gamma$, an analogous equation for $\widehat{F}_\beta({\bf x}/{\bf y}, t)$ is obtained,
\begin{equation}
\partial_t \widehat{F}_{\beta}({\bf x} /{\bf y},t) =
{\mathcal L}_{y,\beta}^+[\widehat{F}_{\beta}({\bf x} /{\bf y},t),\{ \widehat{F}_\alpha\}_{\alpha=1}^N] 
\label{eq3_30}
\end{equation}
which is the relevant equation for practical calculations.

Next, consider the moment hierarchy of the first transit-times
$\{ T_\beta^{(n)}({\bf x}/ {\bf y}) \}_{\beta=1}^N$, $n=0,\dots,N$.
For simplicity, we drop the dependence on the final state ${\bf x}$,
rewriting $T_\beta^{(n)}({\bf x}/ {\bf y})$ simply as $T_\beta^{(n)}({\bf y})$,
\begin{equation}
T_\beta^{(n)}({\bf y}) = \int_0^\infty t^n \, \widehat{F}_\beta({\bf x}/ {\bf y}, t) \, dt 
\label{eq3_31}
\end{equation}
From eq. (\ref{eq3_30})  one obtains the system of equations
for $T_\beta^{(n)}({\bf y})$
\begin{equation}
{\mathcal L}_{y,\beta}^+[T_\beta^{(n)}({\bf y});\{ T_\alpha^{(n)}\}_{\alpha=1}^N] + n \, T_\beta^{(n)}({\bf y}) =0
\label{eq3_32}
\end{equation}
Since $T_\beta^{(0)}({\bf y})=1$, the equations
for the mean first-transit times become
\begin{equation}
{\mathcal L}_{y,\beta}^+[T_\beta^{(1)}({\bf y}),\{T_\alpha^{(1)}\}_{\alpha=1}^N] + 1 =0  
\label{eq3_33}
\end{equation}
and the average mean first-transit time $\langle T^{(1)}({\bf y}) \rangle$
is the weighted average of the $T_\beta^{(1)}({\bf y})$ with respect to the
initial probabilities $\pi_\beta=\mbox{Prob}[\chi_N(0)=\beta]$
\begin{equation}
\langle T^{(1)}({\bf y}) \rangle = \sum_{\beta=1}^N \pi_\beta \, 
T_\beta^{(1)}({\bf y}) 
\label{eq3_34}
\end{equation}
Eqs. (\ref{eq3_32}) are equipped with the boundary condition
$T_\beta^{(n)}({\bf x})=0$ for $n \geq 1$.

\subsection{A simple example}
\label{sec_3_3}

In this paragraph we do not consider any specific physical problem 
involving the estimate of the mean first-transit times, 
but rather a very simple - the simplest - example that highlights the wave-like nature of GPK and the
role of boundary conditions.

Consider  the mean first transit-time at $x=1$ for a pure Poisson-Kac
diffusion problem in $[0,1]$ where the boundary at $x=0$
is impermeable to transport.
The stochastic model is thus
\begin{equation}
d x(t) =  b \, (-1)^{\chi(t)} \, dt
\label{eq3_35}
\end{equation}
where the Poisson process $\chi(t)$ is characterized by
the transition rate $a$.
Let $\alpha=1$ be the state characterized by $(-1)^{\chi(t)}=1$,
and $\alpha=2$ the other state $(-1)^{\chi(t)}=-1$, and use
the notation $T^{(+)}(y)= T_1^{(1)}(y)$, and $T^{(-)}(y)= T_2^{(1)}(y)$.
The equations for $T^{(\pm)}(y)$ in the present case follows
from eq. (\ref{eq3_33})
\begin{eqnarray}
b  \,\frac{d T^{(+)}(y)}{d y} - a \, T^{(+)}(y) + a \, T^{(-)}(y) + 1 & = & 0
\nonumber \\
-b \, \frac{d T^{(-)}(y)}{d y} + a \, T^{(+)}(y) - a \, T^{(-)}(y) + 1 & = & 0
\label{eq3_36}
\end{eqnarray}
equipped with reflecting boundary conditions at $x=0$,
\begin{equation}
T^{(-)}(0)= T^{(+)}(0) 
\label{eq3_37}
\end{equation}
and with the exit conditions at $y=1$,
\begin{equation}
T^{(+)}(1)=0 \, , \qquad T^{(-)}(1)=0 
\label{eq3_38}
\end{equation}
Equations (\ref{eq3_37}) have been derived by several authors
in the one-dimensional case of Poissonian dichotomous noise 
\cite{transit1,transit2,transit3,transit4},
considering also the effect of potential fields and  energy barriers.
Here, we are not interested in the physics of transition rates,
but essentially  on the undulatory mechanism of the propagation of Poisson-Kac dynamics and of its phenomenological implications.

\begin{figure}[h!]
\begin{center} 
{\includegraphics[height=6cm]{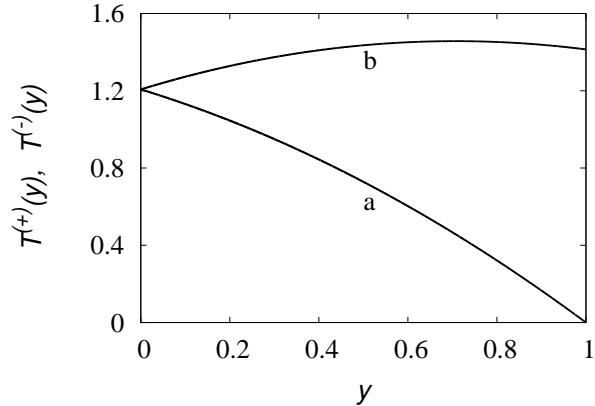}}
\end{center}
\caption{Behavior of $T^{(+)}(y)$ (line a)  and $T^{(-)}(y)$ (line b) vs $y$ 
in the case of the pure Poisson-Kac stochastic process (\ref{eq3_35})
with $a=1$ and $D=1$.
The lines represent the 
solution of eqs. (\ref{eq3_36}) with the boundary conditions
(\ref{eq3_37})-(\ref{eq3_38}).}
\label{Fig2} 
\end{figure}

From the wave-like nature of the process,  solely
the condition for $T^{(+)}(y)$ at $y=1$ propagates, as the
first equation (\ref{eq3_36}) represents a backward 
wave
with respect to the initial position $y$.
Consequently,  $T^{(+)}(y)$ is a continuous function of $y$,
propagating the transit-time condition $T^{(+)}(1)=0$ smoothly
up to the reflecting boundary, located at  $y=0$.  This is not the
case of $T^{(-)}(y)$. As $T^{(-)}(y)$ describes the mean
first-transit time through $y=1$, starting from the initial
state $(-1)^{\chi(0)}=-1$,  in order to reach $y=1$,  a particle must
 travel, first
backwardly in space, and solely after the first switching of the
stochastic Poisson process - that can occur either because of the
internal Poissonian recombination between forwardly and backwardly
directed perturbations, or because of the reflection at $y=0$ -
it can travel in the direction of positive $y$ and reach
the exit point located at $y=1$.
This explains the behavior of $T^{(+)}(y)$
and $T^{(-)}(y)$ depicted in figure \ref{Fig2}.
Focusing attention on $T^{(-)}(y)$,  which is the most interesting quantity,
this function is discontinuous at the exit point $y=1$, as $T^{(-)}(1)=0$,
but the left limit is different from zero, $\lim_{y \rightarrow 1^-}
T^{(-)}(y) \neq 0$.

Moreover, it follows from the above observations,
 that $T^{(-)}(y)$ does not need to be a monotonic
function of $y$. Backwardly oriented particles,
initially placed say at $y_1$, can reach sooner the exit point $y=1$
than particles initially located closer to the exit points, i.e.,
at $y_2>y_1$, due to the recombination/reflection mechanism
described above.

\begin{figure}[h!]
\begin{center}
{\includegraphics[height=6cm]{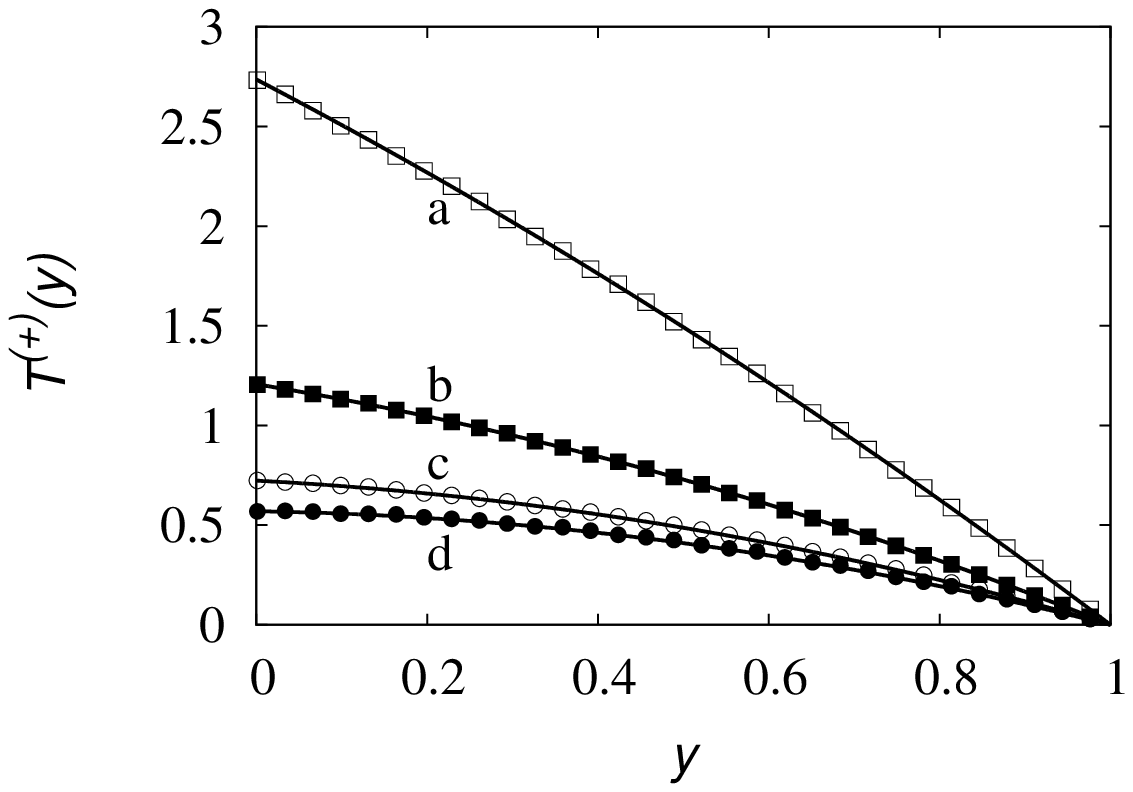}}
\hspace{-1.0cm} {\large (a)} \\
{\includegraphics[height=6cm]{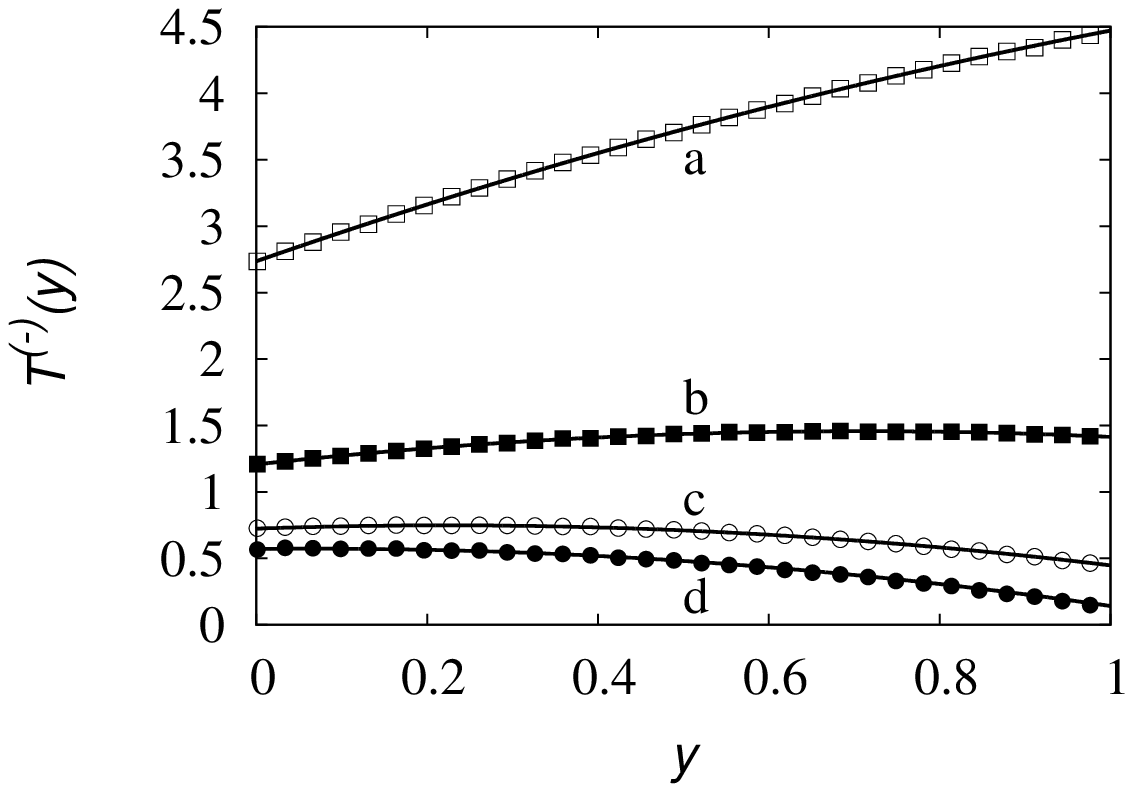}}
\hspace{-1.0cm} {\large (b)}
\end{center}
\caption{Comparison of the solutions of the first transit-time
equations (\ref{eq3_36})-(\ref{eq3_38}) (lines) and
Monte Carlo simulations of the stochastic dynamics (\ref{eq3_35}) (symbols)
at $D=1$, for different values of $a$. Panel (a): $T^{(+)}(y)$ vs $y$,
panel (b) $T^{(-)}(y)$ vs $y$. Line (a) and ($\square$): $a=0.1$;
line (b) and ($\blacksquare$): $a=1$; line (c) and ($\circ$): $a=10$;
line (d) and ($\bullet$): $a=100$.}
\label{Fig3}
\end{figure}

To complete the picture, figure \ref{Fig3} depicts the
comparison of the solution of the first transit-time problem
(\ref{eq3_36})-(\ref{eq3_38}) with Monte-Carlo simulations.
These simulations have been performed integrating eq. (\ref{eq3_36})
with a time step $h_t=10^{-3}$, using an ensemble of $10^5$
particles for each initial position.
Fixing $D=1$, non-monotonicity of $T^{(-)}(y)$ occurs for values of $a$
close to $1$.  For very low values of $a$ (line (a) in panel (b)).
the profile of $T^{(-)}(y)$, is  a monotonically increasing function of $y$,
meaning that the exit through $y=1$ is controlled by reflections at $y=0$.
Conversely, for values of $a\gg 1$, internal recombination effect prevails,
and $d T^{(-)}(y)/dy<0$.

Observe that $T^{(+)}(y)$ and $T^{(-)}(y)$ are in general
radically different from each other, 
so that the estimate of the mean transit time
$\langle T^{(1)}(y) \rangle$ via eq. (\ref{eq3_34}) depends
significantly on the probability weights $\pi_\beta$,
$\beta=1,\dots,N$, i.e., on the state of the noise
perturbation at time $t=0$.

\section{Smoothness of trajectories and emergent fractal properties}
\label{sec_4}

The most remarkable property of Poisson-Kac and GPK processes
is their intrinsic  trajectory smoothness, 
provided that the rates $\{\lambda_\alpha\}_{\alpha=1}^N$ and the norms of the characteristic velocity vectors
$\left\{ \bf b  \right \}_{\alpha=1}^N$ are bounded.  This
means that, with probability $1$, the trajectories of these processes are almost
everywhere differentiable function of time $t$. Differentiability
is lost at any time instant, say $t^*$, when  a transition from  a state $\alpha$ to a state $\beta$  occurs,
but these time instants form a countable
sequence (with probability $1$) and, in any case,  left and right velocities
can be defined at $t^*$, such that
\begin{equation}
\frac{d {\bf x}(t^*_+)}{ d t} - \frac{d {\bf x}(t^*_-)}{ d t} =
{\bf b}_\beta - {\bf b}_\alpha 
\label{eq4_0}
\end{equation}
Smoothness marks a deep distinction with respect to Langevin
equations driven by Wiener processes, which possess
fractal trajectories characterized by a fractal dimension $d_T=3/2$
 \cite{falconer}.

Notwithstanding this basic property, purely diffusive GPK
processes  possess
emergent fractal properties, for any bounded values of the rate
and velocity parameters,  analogous to those characterizing
 the corresponding
Wiener-driven diffusive counterparts.
We define purely diffusive, any GPK process driven exclusively
by unbiased ($\sum_{\alpha=1}^N {\bf b}_\alpha=0$) 
stochastic perturbations, such that the deterministic
component of the dynamics is absent, ${\bf v}({\bf x})=0$,
in eq. (\ref{eq2_8}), and such that a Kac limit exists \cite{giona_gpk}.

This means that a form of ``convergence'' towards  the classical
Brownian behavior can be observed at sufficiently long-times,
for any values of the parameters.

In order to highlight this result using a  physically interesting
example, we analyze the case of a  two-dimensional
Poisson-Kac diffusion in a confined geometry represented
by a deterministic fractal structure.
 Subsequently we analyze the implications of the emergent fractality
 considering a pattern formation problem
arising from colloidal aggregation theory \cite{meakin}.
Poisson-Kac stochastic processes on fractals have been studied
in \cite{giona_bc}. The content of paragraph \ref{sec_4_1}
is therefore a succinct review of the analysis developed
in \cite{giona_bc}, albeit using different fractal models, to
avoid repetition.
Although, some degree of overlapping exists between the
content of paragraph \ref{sec_4_1} and the analysis is \cite{giona_bc},
this review is nonetheless functional to the novel analysis developed
in paragraph \ref{sec_4_2}, and further in \ref{sec_5}.

\subsection{Poisson-Kac diffusion on fractals}
\label{sec_4_1}

Consider the two-dimensional Poisson-Kac model  (\ref{eq2_11}).
In the Kac limit, it converges  to a pure Brownian diffusion
process characterized by a diffusion coefficient $D= b^2/2a$.
Set $D=1$. This means that, for any value of $a$,
the characteristic velocity $b$ attains the  value $b=\sqrt{2 \, a}$.

Instead of analyzing how fractality of Poisson-Kac trajectories
originates in free-space propagation, consider the
same problem in a deterministic two-dimensional fractal structure,
which is physically more appealing.

\begin{figure}[h!]
\begin{center}
\hspace{-1.5cm}
{\includegraphics[height=3.5cm]{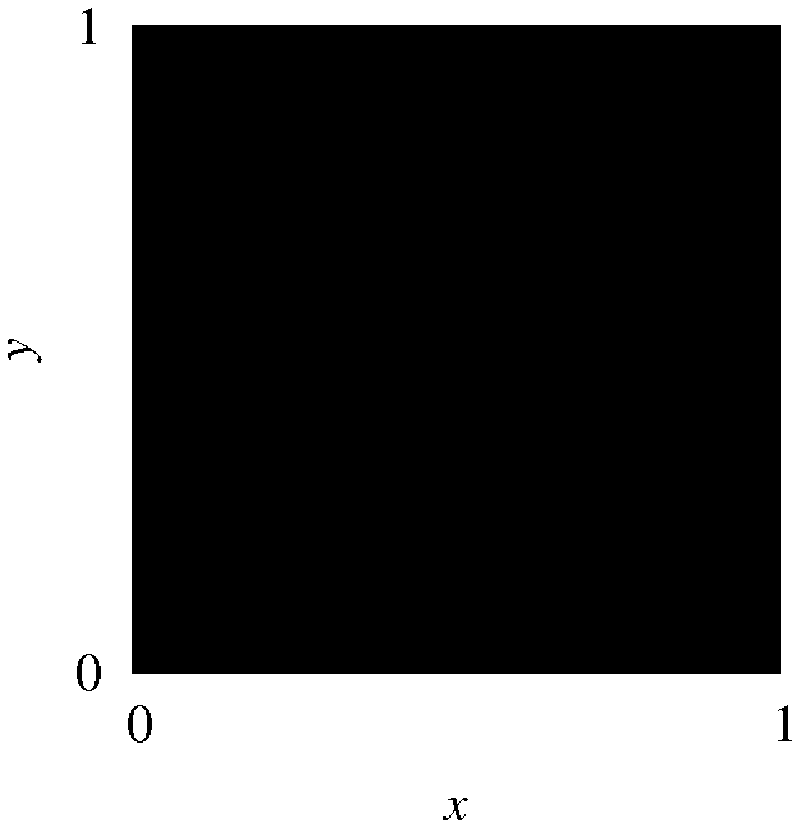}}
\hspace{-0.5cm} {\large (a)} 
{\includegraphics[height=3.5cm]{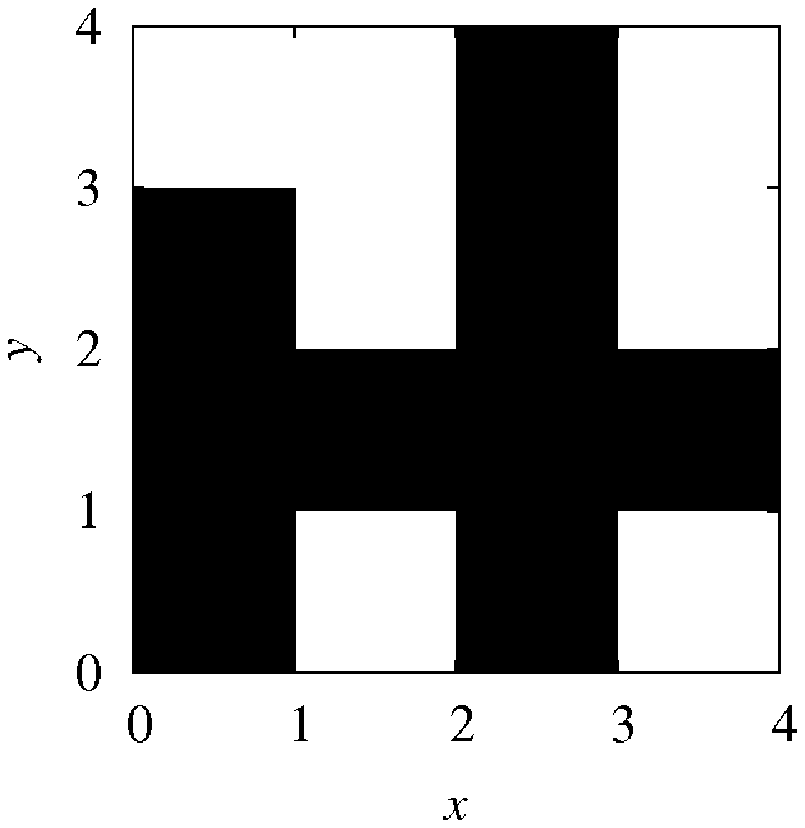}}
\hspace{-0.5cm} {\large (b)}
{\includegraphics[height=3.5cm]{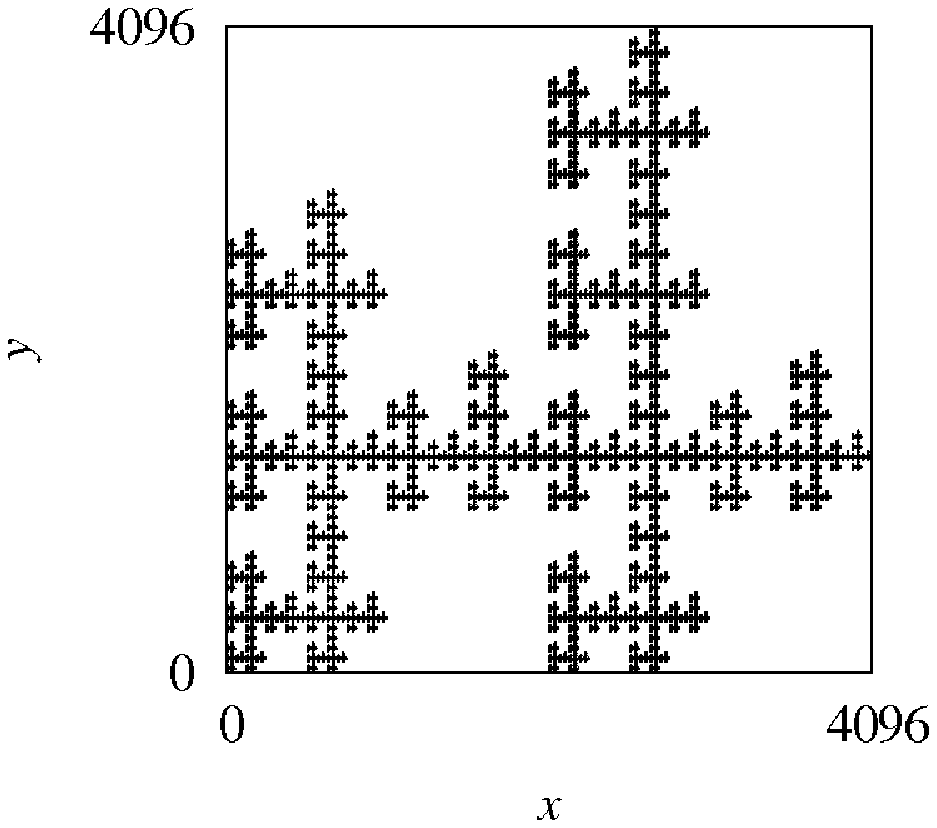}}
\hspace{-0.5cm} {\large (c)}
\end{center}
\caption{Generation process of the loopless deterministic fractal structure
considered. (a) basic unit, (b) first iteration, (c) sixth iteration.}
\label{Fig4}
\end{figure}

The model structure considered is depicted in figure \ref{Fig4}. It is
a loopless fractal that can be generated using $N_t=9$ affine
transformations rescaling the length by a factor of $4$.
Its fractal dimension is therefore, $d_f= \log 9 / \log 4 \simeq 1.584$.

Brownian diffusion in fractal structure is anomalous \cite{havlin,havlin1}, 
meaning that the
mean square displacement $\langle r^2(t) \rangle$ of Brownian particles
scales with time $t$ as
\begin{equation}
\langle r^2(t) \rangle  \sim t^{2/d_w} 
\label{eq4_1}
\end{equation}
where $d_w>2$ is the walk dimension. Since the fractal structure
considered is loopless and possesses a chemical dimension
$d_{C}=1$ \cite{havlin}, the classical theory of transport on fractals
provides for $d_w$ the expression \cite{havlin,havlin1}
\begin{equation}
d_w = d_f+1 \simeq 2.584 
\label{eq4_2}
\end{equation}
Consider the Poisson-Kac diffusion process (\ref{eq2_11})
in this structure. To be precise, the structure considered
is depicted in figure \ref{Fig4} (c), and consists of the
sixth-iteration pre-fractal in the construction process.
We assume that the elementary unit of this structure possesses
unit size (figure \ref{Fig4} (a)), so that
the sixth iteration possesses a maximum length size $L_{\rm max}^{(6)}=4096$,
as depicted in figure \ref{Fig4} (c).

\begin{figure}[h!]
\begin{center}
{\includegraphics[height=7cm]{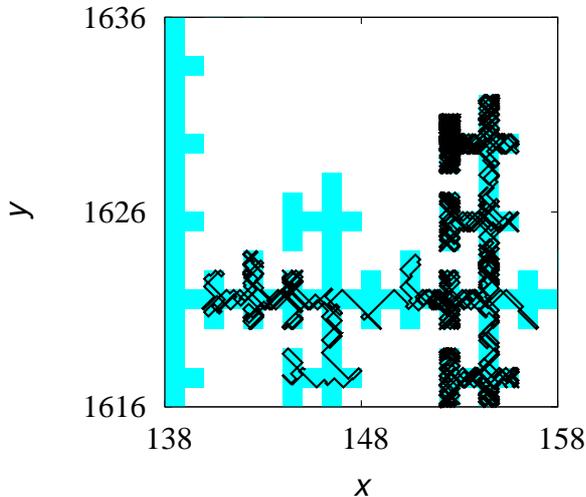}}
\end{center}
\caption{Portion of the orbit of a Poisson-Kac particle diffusing
in the fractal structure considered at $a=1$, $D=1$.}
\label{Fig5}
\end{figure}

Poisson-Kac diffusion is simulated in an off-lattice way, by integrating
eq. (\ref{eq2_11}) starting from  a point chosen at random inside the fractal
structure. Reflection conditions are assumed at the boundary of the
fractal set.
Figure \ref{Fig5} depicts  a portion of the
orbit of a Poisson-Kac particle at $a=1$. The orbit
is almost everywhere smooth, and discontinuity in the
velocity arise as a consequences either  of the internal Poissonian
switching or of the reflection at the boundary \cite{giona_bc}.
At longer time-scales, this regularity is broken as a consequence
of the two mechanisms mentioned above, and the anomalous features
of Brownian motion
can be recovered as a long-term property.

Instead of performing the classical analysis of the mean square displacement,
consider a trajectory-based approach, consisting
in characterizing the anomalous transport properties
using the length-resolution analysis of Poisson-Kac trajectories.
Take the trajectory  of a Poisson-Kac particle, obtained
by integrating eq. (\ref{eq2_11}) from time $t=0$
up to $t=t_{\rm max}$. We use $h_t$ as the integration step
of eq. (\ref{eq2_11}), and $t_{\rm max}= h_t \times (4096)^2$.
Let $L(\Delta t)$ be the length of this portion of a trajectory
estimated by sampling it with a sampling time $\Delta t \geq h_t$.
For a fractal curve \cite{tricot}
\begin{equation}
L(\Delta t) \sim \Delta t^{1-d_T} = \Delta t^{H-1} \,  ,
\label{eq4_3}
\end{equation}
where $d_T$ is the trajectory dimension and $H$ its H\"{o}lder exponent,
related to $d_T$ by the equation $d_T=2-H$.

Since $H=1/d_w$,   it is possible to estimate
the walk dimension $d_w$ (and consequently the basic anomalous features
of transport in fractals) enforcing the length-resolution
analysis, using the relation
\begin{equation}
L(\Delta t) \sim \Delta t ^{1/d_w-1} 
\label{eq4_4}
\end{equation}

\begin{figure}[h!]
\begin{center}
{\includegraphics[height=7cm]{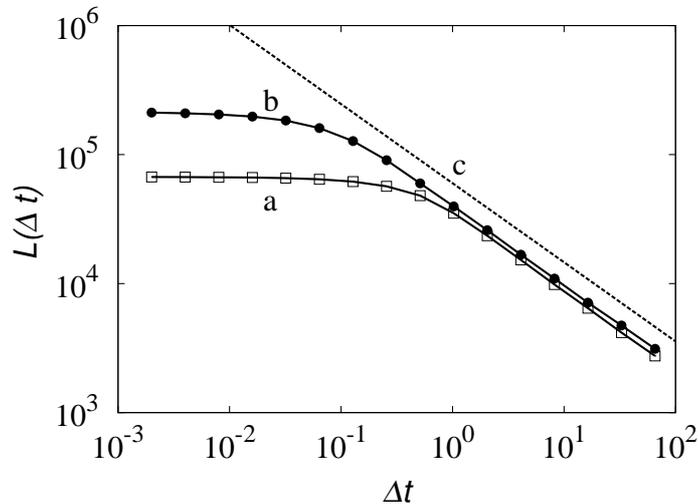}}
\end{center}
\caption{ Graph of the length $L(\Delta t)$ vs temporal yardstick $\Delta t$ 
for a trajectory of a Poisson-Kac particle diffusing  in the loopless
deterministic fractal structure considered at $D=1$. Line (a) 
and ($\square$): $a=1$; line (b) and ($\bullet$): $a=10$.
Dashed line (c) represents the scaling $L(\Delta t) \sim \Delta t^{1-1/d_w}$
with $d_w\simeq 2.584$.}
\label{Fig6}
\end{figure}

Figure \ref{Fig6} depicts the results of this analysis for
two values of $a=1,\,10$. As expect, at higher resolutions (small
$\Delta t$), $L(\Delta t) \rightarrow \mbox{constant}$, due to
the regularity of the trajectories (implying $d_T=1$, and 
ballistic motion $d_w=1$). Above a crossover  resolution $\Delta t_c$,
which depends on $a$, the fractal scaling eq. (\ref{eq4_4}) appears,
and the value of $d_w$ predicted by eq. (\ref{eq4_4})
nicely agrees with the theoretical
estimate of $d_w$  given by eq. (\ref{eq4_2}), namely
$1-1/d_w \simeq 0.613$.

To conclude, even in the presence of complex
geometrical constraints, Poisson-Kac processes equipped
with reflecting boundary conditions 
approach in the long-term  limit the characteristic fractal behavior
typical of Brownian particles diffusing in the same structures.

In the cases of Poisson-Kac processes, fractality is not an
intrinsic feature of the stochastic motion, but  rather a long-term
emergent feature that appears as a consequence of the internal
recombination amongst the states $\alpha=1,\dots,N$, and of the
boundary reflections.
In the case of fractal structures,  boundary reflections primarily
 modulate the fractality of Poisson-Kac trajectories, causing
the modification of the emergent fractal dimension
 from $d_T=3/2$ (as it would
occur in the free space or in bounded Euclidean domains) to
$d_T= 2- 1/d_w > 3/2$.

Fractality as an emergent feature, and not as an intrinsic property
of the fluctuations, modifies the way of considering
physical properties that depends  locally on time, but leave
unmodified, with respect to the classical Brownian
paradigm, the physical manifestations of  long-term
diffusional properties. These two issues are addressed in
the next paragraph, that analyzes fluctuation-controlled
pattern formation phenomena, and further in Section \ref{sec_5},
that addresses the thermodynamics of GPK processes.

\subsection{Emerging fractal properties: the case of DLA clusters}
\label{sec_4_2}

A first implication of the analysis developed above
is that all the processes controlled by  the long-term  dynamics
of the stochastic fluctuations possess the same qualitative
and quantitative properties
 in  the  Poisson-Kac, and in the classical
Brownian case (corresponding to the Kac limit
of Poisson-Kac fluctuations). Therefore, in the case of purely stochastic
motion (i.e. whenever deterministic fields are not superimposed),
Poisson-Kac processes are ``emergently Brownian''.
Let us illustrate this issue via a classical problem controlled
by diffusion: the formation of clusters and aggregates when
the limiting step is diffusion, process known as
Diffusion  Limited Aggregation (DLA) \cite{meakin}.
It is well known that, whenever diffusion is the rate limiting step,
particle aggregates display fractal properties. This process
can be modeled in a very simple way on a lattice, using
the well known DLA algorithm. In its lattice version, 
a seed particle is placed in a middle of the lattice, say
a two-dimensional square lattice. Here ``square'' indicates the topology of
nearest neighboring site: for a site $(h,k)$, $h,k \in {\mathbb N}$, its nearest
neighbors are four, $(h \pm 1,k)$, $(h,k \pm 1)$.
Far away from the seed, a new particle is randomly placed and diffuses
with uncorrelated and independent increments (lattice Brownian motion).
Once it reaches one of the neighboring site of a site belonging to the
aggregate it sticks to it, increasing the cluster structure. A new particle
is then launched and the  process continues.

Instead of Brownian diffusion, Poisson-Kac processes can be
used to si\-mulate the stochastic motion of the particles.
Specifically, we use the following recipe in the simulations:
(i) Each particle possesses a unit size; (ii) particles move according
to eq. (\ref{eq2_11}) i.e., in a continuous off-lattice way;
(iii) any new particle is initially placed far away from the growing
cluster: in the simulations, we locate initially a new particle
at random at a distance of $d_{\rm max}$ (equal to 500 lattice units in the
simulations) from the
closest site of the aggregate; (iv) if a particle escapes away,
i.e., it reaches a maximum distance $d_{\rm ext}=2000$ lattice units from
the cluster, it is discarded, ad a new particle launched;
(v) if it reaches one of  the nearest-neighboring sites of
a cluster site it aggregates; (vi) the process continues
until $N_p=10^5$ particles have formed a connected cluster.

It should  be noted, that the algorithm adopts a continuous
description of the  stochastic particle  motion, while
it keeps a lattice representation of the aggregation
process, based on the topology of a square lattice.

As  regards cluster structure and formation, the
long-term properties of the stochastic motion matter.
Consequently, we expect that  no
significant  differences should appear in the Poisson-Kac case 
with respect to
the classical Brownian-motion case, 
for any values of $a$ (the other parameter $b$
is fixed by the assignment of the effective diffusivity $D=1$).

\begin{figure}[h!] 
\begin{center}
{\includegraphics[height=5cm]{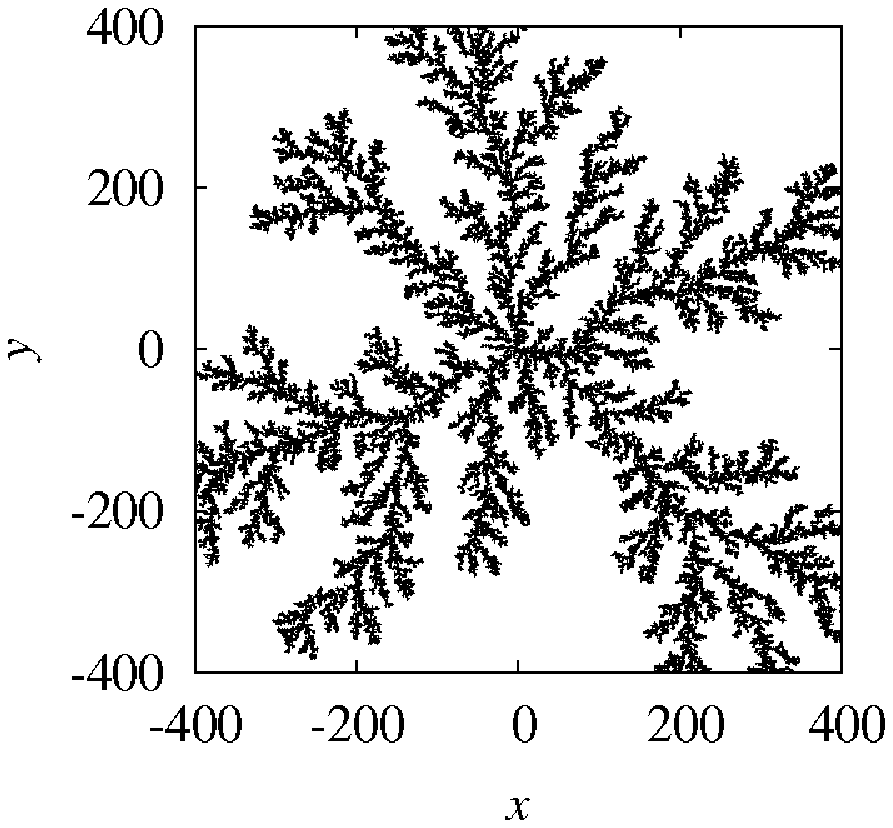}}
\hspace{-1cm} {\large (a)} 
{\includegraphics[height=5cm]{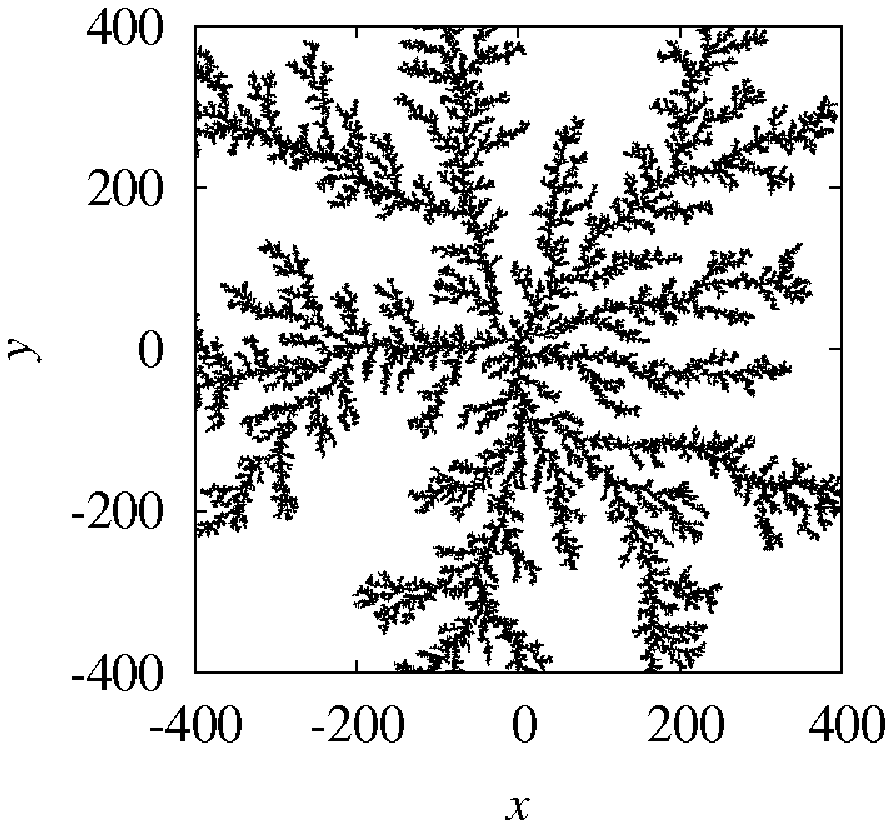}}
\hspace{-1cm} {\large (b)}
\end{center}
\caption{Portion of two-dimensional DLA clusters obtained using 
an off-lattice Poisson-Kac  diffusion process at $D=1$.
Panel (a): $a=1$; panel (b): $a=10$. The seed particle
is initially located at $(0,0)$.}
\label{Fig7}
\end{figure}

Figure \ref{Fig7} depicts the central portion of two aggregates,
obtained using the algorithm described above at two different values
of $a$.
 
\begin{figure}[h!]
\begin{center}
{\includegraphics[height=7cm]{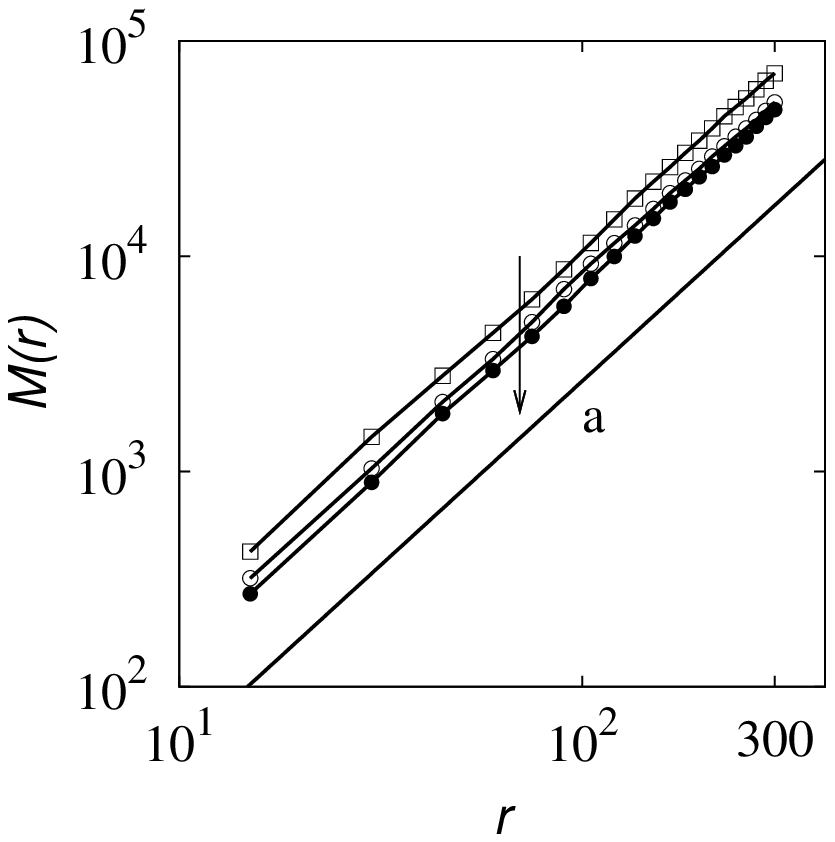}}
\end{center}
\caption{Mass-radius scaling of DLA clusters obtained via Poisson-Kac
diffusion at $D=1$. The arrow indicates increasing values 
of $a$: ($\square$): $a=0.1$; ($\circ$): $a=1$; ($\bullet$): $a=10$.
Line (a) represents the scaling $M(r) \sim r^{d_f}$ with $d_f=1.71$.}
\label{Fig8}
\end{figure}

No significant qualitative differences with respect to
Brownian DLA clusters (depicted e.g. in  \cite{meakin}) can be observed.
A  quantitative check of this statement can be based
on the mass-radius scaling of the clusters, i.e., on the scaling of the
 mass of the aggregate
$M(r)$ (assuming unit density) enclosed within a
circle of radius $r$ centered at the initial seed-particle location.
The fractal nature of the aggregate dictates that
\begin{equation}
M(r) \sim r^{d_f}
\label{eq4_5}
\end{equation}
where the fractal dimension for two-dimensional DLA clusters
is known to be $d_f=1.71 \pm 0.05$.
Figure \ref{Fig8} depicts the results of the mass-radius
scaling of Poisson-Kac DLA clusters at $D=1$, for three values
of the rate $a$.

As expected, Poisson-Kac clusters admit the same fractal scaling
properties of their Brownian-motion counterparts, represented by
the solid curve (a) in figure \ref{Fig8}.
As a second-order effect, it can be observed  that
the switching rate influences slightly the prefactor of
the scaling law (\ref{eq4_5}): at lower  values of $a$,  clusters are
slightly
more dense. 

\section{Smoothness and local energetics of GPK processes}
\label{sec_5}

Diffusion-limited growth is a physical example in which 
long-term properties of fluctuations matter. On the opposite side, whenever
local, short-term properties count, GPK processes possess
a completely different physics and statistical characterization
than their Brownian-motion counterparts. This difference is
ultimately related to the  almost-everywhere local differentiability
of Poisson-Kac trajectories.

This is the case of the energetic theory of GPK 
processes, that  is briefly outlined  below, just to
connect it with trajectory regularity.
The energetics of one-dimensional Poisson-Kac processes 
is developed in \cite{giona5}.

Consider a Ornstein-Uhlenbeck process expressed in the
form of a Generalized Poisson-Kac equation
for a particle of mass $m$ in the presence of a conservative
field $-\nabla_x \Phi({\bf x})$ ($\nabla_x$ is the nabla operator
with respect to the spatial coordinates), and of a  friction force
$-\eta \, {\bf v}$,
\begin{eqnarray}
d {\bf x}(t) & = & {\bf v}(t) \, dt 
\label{eq5_1} \\
m \, d {\bf v}(t) & = & - \nabla_x \Phi({\bf x}(t)) \, d t
- \eta \, {\bf v}(t) \, dt + {\bf b}(\chi_N(t)) \, d t 
 \nonumber 
\end{eqnarray}
In eq. (\ref{eq5_1}) the vectors ${\bf b}_\alpha$ characterizing
the GPK process have the physical dimension of an acceleration.
Taking the scalar product with ${\bf v}(t)$ in the
equation of motion (\ref{eq5_1}),  the energy balance follows
\begin{equation}
d E_{\rm kin}(t)= d L_{\Phi}(t) -d Q(t)+ d L_{\rm stoca}(t)
\label{eq5_2}
\end{equation}
where
\begin{equation}
d E_{\rm kin}(t)=   d \left (  m \, \frac{|{\bf v}(t)|^2}{2} \right ) 
\label{eq5_3}
\end{equation}
is the differential of the kinetic energy,
\begin{equation}
d L_{\Phi}(t) = - \nabla_x \Phi({\bf x}(t)) \cdot {\bf v}(t) \, dt   
\label{eq5_4}
\end{equation}
is the work exerted  by the conservative forces in the interval $[t,t+dt)$,
\begin{equation}
d Q(t)= \eta \, |{\bf v}(t)|^2 \, dt 
\label{eq5_5}
\end{equation}
is the heat released by friction in $[t,t+dt)$, and
\begin{equation}
d L_{\rm stoca}(t)  = {\bf b}(\chi_N(t)) \cdot {\bf v}(t) \, d t 
\label{eq5_6}
\end{equation}
is the stochastic work in $[t,t+dt)$, performed by the stochastic
perturbation.
Observe that, in dealing with GPK processes, there is no
ambiguity in the definition of the stochastic integrals
which characterizes the Wiener case, and induces to assume,
for consistency with the classical expression of the
 kinetic energy, a Stratonovich formulation of the stochastic
integration rule.
As discussed in \cite{giona5}, this ambiguity stems ultimately from
the fractal nature of the  Wiener fluctuations.
Since  stochastic processes driven by  GPK fluctuations
are a.e. smooth, the  stochastic integral for  $L_{\rm stoca}(t)$
  can be viewed simply as an ordinary
Riemann integral.

Since ${\bf b}(\chi_N(t))$ and ${\bf v}(t)$ are bounded, it is
possible to introduce for GPK processes the notion
of {\em local power delivered by the stochastic perturbation} $\pi(t)$,
by taking the time-derivative of $L_{\rm stoca}(t)$,
\begin{equation}
\pi(t) = \frac{d L_{\rm stoca}(t)}{d t}
\label{eq5_7}
\end{equation}
From eq. (\ref{eq5_6}), it follows that
\begin{equation}
\pi(t)=  {\bf b}(\chi_N(t)) \cdot {\bf v}(t) 
\label{eq5_8}
\end{equation}
This concept has been  introduced in \cite{giona5} for one-dimensional
Poisson-Kac processes.
The stochastic power defined by eq. (\ref{eq5_8}) is
itself a stochastic
process. Let $F_\pi(\pi,t)$
be its distribution function and $p_\pi(\pi,t)$ its
probability density function.
Since ${\bf b}(\chi_N(t))$ can attain solely
$N$ distinct values ${\bf b}_\alpha$, $\alpha=1,\dots,N$,
the distribution function $F_\pi(\pi,t)$ can be
expressed in terms of the partial probability waves
$\overline{p}_\alpha({\bf x},{\bf v},t)$ as
\begin{equation}
F_\pi(\pi,t) = \sum_{\alpha=1}^N \int_{ {\mathbb R}^3 } d {\bf x}
\int_{\{ {\bf b}_\alpha \cdot {\bf v} < \pi \}}
\overline{p}_\alpha({\bf x}, {\bf v}, t) \, d {\bf v} 
\label{eq5_9}
\end{equation}
where the integral over ${\bf x}$ is extended over the entire
spatial domain ${\mathbb R}^3$, and the integral
over the velocity over the subset such that ${\bf b}_\alpha \cdot {\bf v} 
< \pi$.
Let ${\bf b}_\alpha=(b_{\alpha,1},b_{\alpha,2},b_{\alpha,3})$
 and ${\bf v}=(v_1,v_2,v_3)$ be the Cartesian representations
of these two vector. Assume, for simplicity,
 that, for any $\alpha=1,\dots,N$,
$b_{\alpha,3} \neq 0$. It follows  from eq. (\ref{eq5_9})  that
\begin{equation}
p_\pi(\pi,t)= \sum_{\alpha=1}^N  \int_{ {\mathbb R}^3 } d {\bf x}
\int_{-\infty}^\infty d v_1 \int_{-\infty}^\infty d v_2 \, 
\overline{p}_\alpha(x_1,x_2,x_3,v_1,v_2, (\pi-b_{\alpha,1} v_1
- b_{\alpha,2} v_2 )/b_{\alpha,3}, t) 
\label{eq5_10}
\end{equation}
which is a closed-form expression for $p(\pi,t)$
relating it to the partial probability densities
 $\overline{p}_\alpha({\bf x},{\bf v},t)$ of the GPK process.

\section{Completeness of the  GPK description}
\label{sec_6}

In this Section we discuss a simple but interesting property
that distinguishes Poisson-Kac and GPK processes from their
Wiener-driven counterparts. This property will be
referred to as {\em completeness of the stochastic description},
for reasons that will be soon clear, and  admits several
thermodynamic implications.

Consider two stochastic models (for notational simplicity
in one spatial coordinate):
a Langevin equation
\begin{equation}
d x(t) = v(x(t)) \, d t + \sqrt{2 \, D} \, d w(t) 
\label{eq6_1}
\end{equation}
where $d w(t)$ are the increments of a one-dimensional
Wiener process, and its Poisson-Kac counterpart
\begin{equation}
d y(t)= v(y(t)) \, dt + b \, (-1)^{\chi(t)} \, dt 
\label{eq6_2}
\end{equation}
where $\chi(t)$ is  a Poisson process  characterized
by a switching rate $a$. In eqs. (\ref{eq6_1})-(\ref{eq6_2}),
$v(x)$ is the same deterministic velocity field,
and $b = \sqrt{2 \, D \, a}$, so that in the Kac limit
eq. (\ref{eq6_2}) converges to eq.  (\ref{eq6_1}).

We can regard eqs. (\ref{eq6_1})-(\ref{eq6_2}) as
the dynamics under overdamped conditions of a particle
in a deterministic field connected to
a heat bath that stochastically supplies energy to it.
The bath is here described in a very ``crude'' way:
as $\sqrt{ 2 \, D} \, d w(t)$ in eq. (\ref{eq6_1}) and
$b \, (-1)^{\chi(t)} \, dt$ in eq. (\ref{eq6_2}), respectively.
It is a very simplified description with respect to
normal-mode characterization \cite{normalmode1,normalmode2}, but it suffices for the present purposes. 

The statistical description of eq. (\ref {eq6_1}) leads to
a Fokker-Planck equation for
the probability density function $p(x,t)$,
\begin{equation}
\partial_t p(x,t) = - \partial_x \left [ v(x) \, p(x,t) \right  ]
+ D \, \partial_x^2 p(x,t) 
\label{eq6_3}
\end{equation}
while  the evolution of the 
two partial probability
waves $p^+(y,t)$ and $p^-(y,t)$
associated with eq. (\ref{eq6_2})  is given by
\begin{eqnarray}
\partial_t p^+(y,t) & = & -\partial_y \left [(v(y)+b) \, p^+(y,t) 
\right ] - a \, p^+(y,t)+ a \, p^-(y,t)
\nonumber \\
\partial_t p^-(y,t) & = & -\partial_y \left [(v(y)-b) \, p^+(y,t) 
\right ] + a \, p^+(y,t)- a \, p^-(y,t) 
\label{eq6_4}
\end{eqnarray}
converging, in the Kac limit, to eq. (\ref{eq6_3})
as  regards the overall probability density function
$p(y,t)=p^+(y,t)+p^-(y,t)$, substituting $y$ with $x$.

Albeit the higher order (second) derivative $\partial_x^2$, 
the Fokker-Planck equation
(\ref{eq6_3}) is considered
 to provide a ``simpler'' and more compact description
of the stochastic dynamics, as for eq. (\ref{eq6_2}) the
statistical characterization of the process
necessitates two partial probability  density functions $p^+(y,t)$,
$p^-(y,t)$ instead of one \cite{dunkelrel}.

This ``complexity counterargument'' has been mentioned in \cite{dunkelrel}
(see also the references therein),
as a practical (computational) disadvantage in the use
of Poisson-Kac processes.

The simplicity and compactness of the Fokker-Planck
formulation (\ref{eq6_3}) that requires a single probability density $p(x,t)$
is the major strength of the classical Langevin model, but at the
same time also an issue.
The statistical structure of Wiener processes, (stemming
from a large-number ansatz in modeling the stochastic
perturbations) makes it possible to renormalize
completely the fine structure of noise
in the associated Fokker-Planck equation.
No information on the state of the heat bath is present
anymore in eq. (\ref{eq6_3}), other
than its ``effective strength'' expressed by the diffusivity
$D$.
The complete renormalization of the stochastic perturbation is
the key essence of Wiener-driven Langevin equations
that ultimately leads to a second-order (parabolic) Fokker-Planck equation.
Eq. (\ref{eq6_1}) describes a stochastically non-isolated system -
the system interacts with the heat bath -
but information on the state of the stochastic surrounding is
completely lost, just because of the 
uncorrelated nature of the increments of a Wiener process and
of their Gaussian distribution.

Completely different is the case of Poisson-Kac dynamics.
The partial probabilities $p^+(y,t)$ and $p^-(y,t)$ account
not only for the state of the system but also for the
local state of the heat bath at time $t$.
The connection between the particle state and  bath conditions
is kept at all the times $t$, so that at any time we can easily
obtain information about their mutual
correlation properties.

The apparent additional complexity of the Poisson-Kac model
turns into a more detailed
and comprehensive description of the interactions of
a physical system under investigation with its
stochastic surrounding.
The memory effects associated with Poisson-Kac processes
are just remnant memory on the state of the stochastic
surrounding, that significantly lead to
the regularity of the trajectories.
In this conceptual perspective, the fractal nature of
Wiener fluctuations, and of system observables driven by
Wiener fluctuations, is just the consequence of a deliberate lack
of memory on the state of the heat bath interacting with
the system under investigation.

We refer  to this property characterizing
 Poisson-Kac  and GPK processes as
``{\em completeness of the stochastic description}'', meaning
with that, for a given stochastic model, the statistical
description of the system (in our case  the partial probability 
waves $p^+(y,t)$,
$p^-(y,t)$) provides a complete statistical
characterization not only of the system, but also of its local
stochastic environment, with which the system  interacts.

As a result of the  completeness in the characterization
of Poisson-Kac and
GPK processes, it is rather  intuitive  to expect that the Markovian
nature of the model should
be expressed via an extended Markov condition
rather than with the strict Markovian one.
All these  arguments applies on equal footing to
GPK processes possessing $N$ states.

The completeness in the stochastic description of a Poisson-Kac and GPK 
 processes is in some sense related to
the regularity of the trajectories, just
because the stochastic bath has a finite  memory of its state.
Completeness permits a simpler and powerful
characterization of the thermodynamics of these
systems. Eqs.  (\ref{eq5_9}) and (\ref{eq5_10}),
relating the statistical characterization to the local stochastic 
power delivered by the heat bath, provide
a first example of this detailed energetic characterization.
 
\section{Some remarks on correlation properties and tunneling}
\label{sec_7}

One of the reasons  for the application of  Poisson-Kac processes is
that they provide a tractable model for colored noise \cite{col1,col2},
as the stochastic perturbation $(-1)^{\chi(t)}$ possesses
an exponentially decaying correlation function \cite{kac}
\begin{equation}
\left \langle (-1)^{\chi(t+\tau)} \, (-1)^{\chi(t)} \right \rangle
=  e^{-2 \, a  |\tau|} 
\label{eq7_1}
\end{equation}
for  all $\tau \in {\mathbb R}$.

A deeper investigation  reveals that  it is its
trajectory regularity, rather than the colored correlation properties
expressed by eq. (\ref{eq7_1}),  the major responsible
for the long-term qualitative  statistical properties in the
presence of deterministic biasing fields superimposed to
stochastic perturbations.
This Section addresses this issue via a simple but highlighting example. 

Consider
a simple one-dimensional Poisson-Kac process in the
presence of a harmonic potential under overdamped
conditions
\begin{equation}
d y(t) = - y(t) \, dt + b \, (-1)^{\chi(t)} \, dt 
\label{eq7_2}
\end{equation}
where the harmonic contribution has been set equal to $- y$ upon
a suitable rescaling of the time variable.

The corresponding Langevin model is given by
\begin{equation}
d x(t) = - x(t) \, d t + d w_{\rm corr}(t) 
\label{eq7_3}
\end{equation}
where $d w_{\rm corr}(t)$ are the increments of a stochastic
process possessing the same correlation properties
as $(-1)^{\chi(t)}$. Such a process can be constructed using a
Markovian embedding $w_{\rm corr}(t)=z(t)$, where
\begin{equation}
d w_{\rm corr}(t)= d z(t) = - 2 \, a \, z(t) \, dt +
\sqrt{2 \, D} \, d w(t) 
\label{eq7_4}
\end{equation}
$w(t)$ being an one-dimensional Wiener process, and the
parameter $D$ is chosen such that $b^2/2 a = D$.

The corresponding Langevin model driven by Wiener fluctuations
is therefore a vector-valued stochastic model ${\bf X}(t)=(X(t),Z(t))$,
${\bf x}=(x,z)$,  described by the equation
\begin{equation}
d {\bf x}(t) = {\bf B} \, {\bf x}(t) + \sqrt{2 \, D} \, {\bf m}
\, d w(t) 
\label{eq7_5}
\end{equation}
where 
\begin{equation}
{\bf B}=
\left (
\begin{array}{ll}
-1 & - 2 a \\
0 & - 2 a
\end{array}
\right )
\, ,
\qquad
{\bf m}
= \left (
\begin{array}{l}
1 \\
1
\end{array}
\right ) 
\label{eq7_6}
\end{equation}
As regards the $x$- $y$-dynamics, eqs. (\ref{eq7_2}) and (\ref{eq7_6})
represent the dynamics of two harmonic oscillators of equal strength,
subjected to two random perturbations possessing identical
correlation properties.
As addressed below, these two models are characterized
by  completely different long-term properties.

Eqs. (\ref{eq7_2}) possesses a unique stationary
invariant measure with a compact support. This
can be easily verified by considering
the dynamics of the two associated
partial probability waves
\begin{eqnarray}
\partial_t p^+(y,t) & = & - \partial_y \left [ f_+(y) \, p^+(y,t) \right ]
- a \, p^+(y,t) + a \, p^-(y,t) \nonumber \\
\partial_t p^-(y,t) & = & - \partial_y \left [ f_-(y) \, p^-(y,t) \right ]
+ a \, p^+(y,t) - a \, p^-(y,t) 
\label{eq7_7}
\end{eqnarray}
where
\begin{equation}
f_+(y)=-y+b \, , \qquad f_-(y)=-y-b
\label{eq7_8}
\end{equation}
Since $p^+(y,t)$ represents a wave propagating forward in the positive
$y$-direc\-tion, and $p^-(y,t)$ a wave moving backwardly, it
is easy to recognize from eq. (\ref{eq7_8}) that the two limit
points $y=\pm b$ represent a form of perm-selective
membrane to transport: at $y=b$ solely the backwardly-oriented
wave $p^-(y,t)$ can cross this point while the forward wave is stopped,
and the opposite occurs at $y=-b$, where $p^+(y,t)$ can propagate
forwardly while it  is a stagnation point for $p^-(y,t)$.

\begin{figure}[h!]
\begin{center}
{\includegraphics[height=7cm]{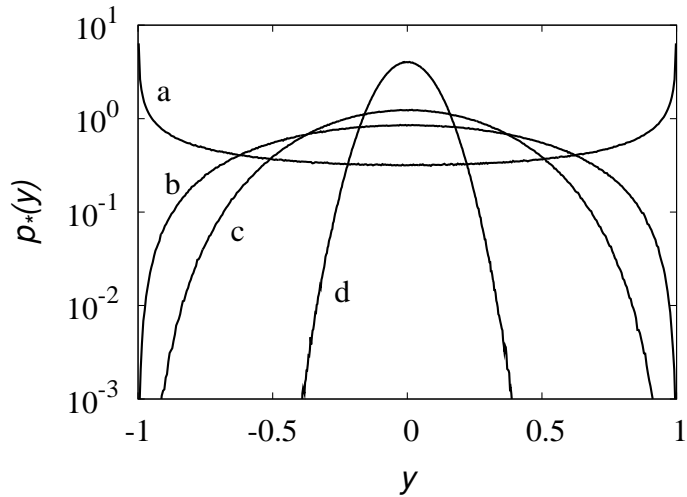}}
\end{center}
\caption{Unique stationary invariant probability density function
$p_*(y)$ vs $y$ for the model (\ref{eq7_2}) at $b=1$ for different
values of $D$. Line (a) $D=1$, line (b) $D=5$, line (c) $D=10$,
line (d) $D=100$.}
\label{Fig9}
\end{figure}

The perm-selective nature of these two points determines that
in the long-term  $(t \rightarrow \infty$) a
unique system of stationary partial waves $p_*^+(y)$, $p_*^-(y)$ establishes,
possessing compact support in the interval $[-b,b]$.
Consequently, there exists a unique stationary probability density
function $p_*(y)=p_*^+(y)+p_*^-(y)$, compactly supported in $[-b,b]$, and vanishing
elsewhere.
This phenomenon is shown in figure \ref{Fig9},
where  several stationary densities $p_*(y)$ are depicted, 
keeping fixed $b=1$, and varying the diffusivity $D$. Data have been obtained 
using an ensemble of $N_p=10^7$ particles moving according to
eq. (\ref{eq7_2}), initially distributed at random, plotting
their long-term stationary density.

Next, consider eqs. (\ref{eq7_5})-(\ref{eq7_6}). The matrix ${\bf B}$
is upper triangular, possessing  eigenvalues $-1$ and $-2 a$.
Let ${\bf T}$  be the transformation matrix with respect to the
eigenvalue basis of ${\bf B}$,
\begin{equation}
{\bf T}=
\left (
\begin{array}{cc}
1 & 1 \\
0 & c 
\end{array}
\right ) \, ,
\qquad c= \frac{2a-1}{2 a} 
\label{eq7_9}
\end{equation}
and introduce the transformed variables
\begin{equation}
{\bf x}^\prime = {\bf T}^{-1} \, {\bf x} = \frac{1}{c}
\left (
\begin{array}{c}
x-z \\
z
\end{array}
\right ) 
= \left (
\begin{array}{c}
x^\prime  \\
z^\prime
\end{array}
\right )
\label{eq7_10}
\end{equation}
In the transformed variables, the dynamics is expressed
 by the equations
\begin{eqnarray}
d x^\prime(t) & = & - x^\prime(t) \, dt + a_1 \, d w(t) 
\nonumber \\
d z^\prime(t)  & = & - 2 \, a \, z^\prime(t) \, dt + a_2 \, d w(t) 
\label{eq7_11}
\end{eqnarray}
where
\begin{equation}
\left (
\begin{array}{c}
a_1 \\
a_2 
\end{array}
\right )
= \sqrt{2 \, D} \, {\bf T}^{-1}
\left (
\begin{array}{c}
1 \\
1 
\end{array}
\right ) 
\qquad
{\bf T}^{-1} = \frac{1}{c}
\, \left (
\begin{array}{cc}
c & -1 \\
0 & 1
\end{array}
\right )
\, .
\label{eq7_12}
\end{equation}

The diagonalization of the coefficient matrix ${\bf B}$ decouples the two 
degrees of freedom. The evolution for $x^\prime(t)$ and $z^\prime(t)$
is that of two decoupled harmonic oscillators in the presence
of Wiener noise. The associated Fokker-Planck
equations can be solved in closed form, but this
is not essential in the present analysis. What is
interesting is that the long-term probability density
functions for  $x^\prime$, $z^\prime$, and for $x$ are
not compactly supported as their support is just  the whole
real line ${\mathbb R}$.

This represents a major qualitative difference between
this model and the Poisson-Kac counterpart.
In point of fact, by choosing  a deterministic biasing field 
$v(y)$ different from that of an elastic force, e.g.  as in \cite{giona3}, it 
is  possible to show that the corresponding eq. (\ref{eq7_2}),
substituting $v(y)$ to $-y$,
is  not even ergodic, but possesses multiplicity of stationary invariant densities, while its Langevin-Wiener counterpart  would admit
a unique stationary invariant density function. The analysis
of this case is addressed in \cite{giona3}, and consequently is not 
repeated here.

We have chosen the simpler example of a harmonic oscillator
in order to highlight a further significant properties of Poisson-Kac
processes strictly connected to their trajectory regularity.

Ascertained from the above analysis that the colored nature of the
stochastic perturbation is not the cause of its peculiar behavior
(compactly supported stationary probability density),
the natural question is to understand its physical-mathematical
origin.

The answer to this question is essentially  related to the  tunneling dynamics 
exhibited in
correspondence to the critical points $y=\pm b$,
where $f_+(y)$, ($y=b$), and $f_-(y)$, ($y=-b$), vanish.
Consider $y=b$, as the analysis of $y=-b$ is perfectly
mirror-symmetric, interchanging the forward wave with the backward one.
The impossibility for the forward wave  (and consequently for
the entire process) to perform a tunneling across $y=b$ towards
values of $y$ greater than $b$ is a consequence exclusively of 
 trajectory regularity. More precisely,
if the stochastic process driving the dynamics would not be smooth,
but it would possess fractal trajectories this tunneling process
would be possible.

\begin{figure}[h!]
\begin{center}
{\includegraphics[height=7cm]{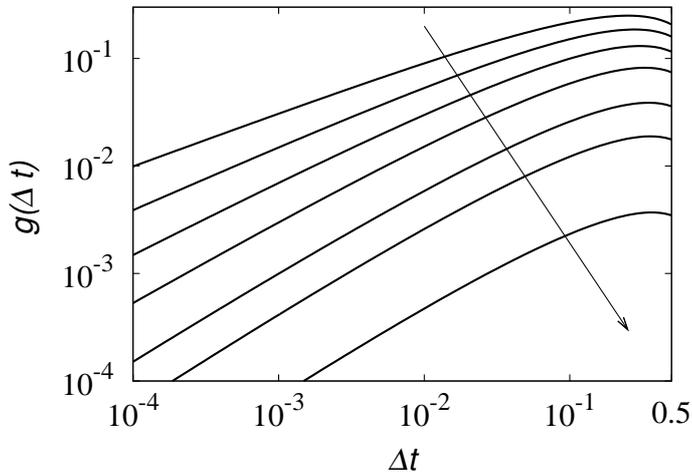}}
\end{center}
\caption{Graph of $g(\Delta t)$ vs $\Delta t$ for several
values of the H\"{o}lder exponent $H$. The direction of the
arrow indicates increasing values of $H=0.5,\, 0.6,\,0.7,\,0.8,\,0.9,\,0.95,\,0.99$. At $H=1$, $g(\Delta t)=0$, identically.}
\label{Fig10}
\end{figure}

This claim can be explained by means of a simple calculation.
Consider a generalized stochastic model of the form
\begin{equation}
d y(t)= v(y(t)) \, dt + a(y(t)) \, d w_g(t) \
\label{eq7_13}
\end{equation}
where $d w_g(t)$ are the increments of some
generic stochastic process (which is specified below), and $a(y)>0$.
Assume that in the neighbourhood of some point $y^*$, say $y^*=0$,
we have $v(y)=-1 + {\mathcal O}(|y-y^*|)$,
$a(y)=1+{\mathcal O}(y-y^*|)$ (we set the values
of $v(y)$ and $a(y)$ at $y^*$ equal to $-1$ and $1$,
just to avoid unnecessary constants), and that in the
neighbourhood of $y^*$ the increments of $d w_g(t)$ behaves
as
\begin{equation}
d w_g(t) = r(t) \, (d t)^H 
\label{eq7_14}
\end{equation}
where $r(t)$ is some random process $|r(t)|\leq 1$.
Any further specification of $r(t)$ is unnecessary, as will
be clear below. Essentially, eq. (\ref{eq7_14})
indicates that the stochastic process $w_g(t)$ is
characterized by a H\"{o}lder exponent $H$.
Let $H\geq 1/2$.
Using a Euler approximation for the dynamics
of $y(t)$ near $y^*$, assuming $y(t)=y^*_-=0_-$, i.e.,  that
$y(t)$ at time
$t$ is arbitrarily close to $y^*=0$, but negative, it
follows that the optimal conditions for tunneling towards
values of $y>y^*$ occur if $r(t)=1$, 
and in this case the Euler approximation provides
\begin{equation}
y(t+\Delta t)-y(t)= -\Delta t + \Delta t^H + o(\Delta t) 
\label{eq7_15}
\end{equation}
From eq. (\ref{eq7_15}), it follows that the
capability for the stochastic process to cross $y^*$, i.e.,
to perform tunneling through it, depends exclusively on
the function
\begin{equation}
g(\Delta t)= - \Delta t+ \Delta t^H 
\label{eq7_16}
\end{equation}
If $g(\Delta t)>0$ for small $\Delta t$, there exists a finite probability
for the particle to pass across $y^*$ towards positive $y$-values.
The function $g(\Delta t)$ depends on the H\"{o}lder exponent $H$
controlling the regularity of the stochastic process.
Figure \ref{Fig10} depicts the behavior of $g(\Delta t)$
vs $\Delta t$ for several values of $H$. 

For any $H<1$
$g(\Delta t)>0$ for small $\Delta t$, corresponding to the occurrence
of tunneling. Conversely, for $H=1$, i.e., for stochastic processes
possessing a.e. smooth  trajectories, 
 $g(\Delta t)=0$ identically, and tunneling
is completely hindered. This is the case of Poisson-Kac processes,
and this simple argument supports the statement that
trajectory regularity and the boundedness of the velocity
(corresponding to the fact that $|r(t)|$ in eq. (\ref{eq7_4})
admits an upper bound) are the two conditions
preventing tunneling effects.

Therefore, tunneling barriers, controlling e.g. the occurrence
of multiple stationary invariant densities in Poisson-Kac model, as discussed
in \cite{giona3}, are the consequence of trajectory regularity which,
in turn, is implied  by  the finite propagation velocity of these 
stochastic perturbations.

This observation opens up interesting observations in quantum
mechanical problems: non-relativistic quantum-particles,
obeying the Schr\"{o}dinger equation,
possess completely different tunneling properties
of their relativistic
counterpart obeying the Dirac equation (see e.g. the Klein paradox
in the relativistic case) \cite{klein1,klein2}.
Since the two quantum models, the non-relativistic Schr\"{o}dinger,
and the relativistic Dirac equations  are related, via analytic
continuation in the time variable, to Wiener and Poisson-Kac processes,
respectively, it is not surprinsing, in the light of what above discussed,
their completely different tunneling
behaviors, which ultimately can be
predicted from their stochastic counterparts. 
The case of tunneling is probably
the indicator of something  more basic:
  the analogy Schr\"{o}dinger/Wiener processes
vs Dirac/Poisson-Kac fluctuations in connection
with  tunneling phenomena 
is just the first point of attack in order to
unveil the deep connections between stochastic
fluctuations and quantum world at a fundamental
physical level, i.e., beyond the simple mathematical
claim that a purely formal analogy exists between these processes.
We hope to analyze this issue in future works.

\section{Concluding remarks}
\label{sec_8}

Starting from the analysis of some basic properties
(Markovian nature, regularity of the trajectories),
we have derived some general results for Poisson-Kac
and GPK processes.

The extended Markovian nature and the notion of stochastic
completeness can be viewed as
two faces of the same coin. If information on the stochastic heat  bath,
perturbing the system, in retained in the statistical
description, it is rather intuitive to expect that
the Markovian nature of the process cannot be
expressed as a strict Markovian condition involving
solely the probability density function $p({\bf x},t)$, but
it should be stated in an extended form using a 
vector-valued system of partial probability densities. This
is the case of GPK processes. The description involves
a finite number $N$ of partial probability densities,
just because the underlying stochastic process (the $N$-state
finite Poisson process) attains solely a finite number of states.
As discussed in \cite{giona6},  if this description is 
generalized to an infinite number of states, then the
corresponding complete stochastic description of the process
 involves a
infinite-dimensional vector of partial probabilities. 

The other relevant issue, thoroughly  addressed in the article, is
the almost everywhere regularity of
the trajectories of GPK processes. The smoothness
of the trajectories influences the thermodynamic formalism,
as it permits to define the notion of  local  power  (time derivative of
work) delivered by the heat bath to the system.

It influences deeply the long-term qualitative properties.
In Section \ref{sec_7}, this qualitative difference between
a Poisson-Kac process and a similar process driven by colored
Wiener fluctuations  has been connected  to the
compactness  of the support of  the unique 
stationary density function.
Using other classes of deterministic biasing fields -
 different from a pure elastic
force - this
difference can be related to  striking 
qualitative differences, such as  ergodicity-breaking and the
occurrence of multiple stationary invariant densities, as discussed
thoroughly in  \cite{giona3}, enforcing a deterministic
bias giving rise to boundary-layer polarization.

As discussed in Section \ref{sec_7}, the compactness of the stationary
probability density in the Poisson-Kac processes is one-to-one
with the tunneling capabilities of these stochastic systems
which, in turn, is related essentially to trajectory
regularity and not on the colored  nature (correlation properties) of these 
stochastic perturbations (which is also a by-product of trajectory
regularity).

The quantum implications of this result, especially
as  regards the relativistic case (the Dirac equation)
will be addressed elsewhere. In point of fact, the relativistic
implications of Poisson-Kac and GPK process   are
essentially based on their trajectory regularity,
i.e., on the fact that their velocities are intrinsically bounded,
 which is the fundamental physical requirement
imposed by the constant value of light velocity {\em in vacuo}.
The use of Poisson-Kac processes in relativistic applications
will be addressed in future works. This is a very delicate
issue involving the meaning of stochastic
fluctuations in relativistic systems and field theories.

\end{document}